\documentclass{article}
\usepackage{jheppub}

\usepackage{ifpdf}

\usepackage{afterpage}

\usepackage{amssymb}
\usepackage{amsmath}
\usepackage{graphicx}
\usepackage{longtable}
\usepackage{verbatim}
\usepackage{amsfonts}

\newcommand{\bear}{\begin{array}}
\newcommand{\ear}{\end{array}}

\newcommand{\beq}{\begin{eqnarray}}
\newcommand{\eeq}{\end{eqnarray}}
\newcommand{\beqa}{\begin{eqnarray}}
\newcommand{\eeqa}{\end{eqnarray}}
\newcommand{\no}{\nonumber}

\def\OMIT#1{{}}
\newcommand{\lsim}{\mathrel{\rlap{\lower4pt\hbox{\hskip1pt$\sim$}}
    \raise1pt\hbox{$<$}}}         
\newcommand{\gsim}{\mathrel{\rlap{\lower4pt\hbox{\hskip1pt$\sim$}}
    \raise1pt\hbox{$>$}}}         

\newcommand{\be}{\begin{equation}}
\newcommand{\ee}{\end{equation}}
\newcommand{\ba}{\begin{eqnarray}}
\newcommand{\ea}{\end{eqnarray}}

\newcommand{\FF}{\mathcal{F}}

\newcommand{\WW}{\mathcal{W}}

\newcommand{\OO}{\mathcal{O}}

\def\lsim{\mathrel{\rlap{\lower4pt\hbox{\hskip1pt$\sim$}}
    \raise1pt\hbox{$<$}}}         
\def\gsim{\mathrel{\rlap{\lower4pt\hbox{\hskip1pt$\sim$}}
    \raise1pt\hbox{$>$}}}         

\begin{document}

\vspace*{-30mm}

\title{\boldmath Natural SUSY Predicts: Higgs Couplings}

\author[a]{Kfir Blum,}
\author[a,b,c]{Raffaele Tito D'Agnolo,}
\author[d]{JiJi Fan}
\affiliation[a]{School of Natural Sciences, Institute for Advanced Study, Princeton, NJ 08540}
\affiliation[b]{Scuola Normale Superiore and INFN, Piazza dei Cavalieri 7, 56126, Pisa, Italy}
\affiliation[c]{CERN, European Organization for Nuclear Research, Geneva, Switzerland}
\affiliation[d]{Department of Physics, Princeton University, Princeton, NJ, 08540}
\emailAdd{kblum@ias.edu}
\emailAdd{raffaele.dagnolo@sns.it}
\emailAdd{jijifan@princeton.edu}

\vspace*{1cm}

\abstract{We study Higgs production and decays in the context of natural SUSY, allowing for an extended Higgs sector to account for a 125 GeV lightest Higgs boson. Under broad assumptions, Higgs observables at the LHC depend on at most four free parameters with restricted numerical ranges. 
Two parameters suffice to describe MSSM particle loops. The MSSM loop contribution to the diphoton rate is constrained from above by direct stop and chargino searches and by electroweak precision tests. Naturalness, in particular in demanding that rare $B$ decays remain consistent with experiment without fine-tuned cancellations, provides a lower (upper) bound to the stop contribution to the Higgs-gluon coupling (Higgs mass). 
Two parameters suffice to describe Higgs mixing, even in the presence of loop induced non-holomorphic Yukawa couplings. Generic classes of MSSM extensions, that address the fine-tuning problem, predict sizable modifications to the effective bottom Yukawa $y_b$. Non-decoupling gauge extensions enhance $y_b$, while a heavy SM singlet reduces $y_b$. 
A factor of 4-6 enhancement in the diphoton rate at the LHC, compared to the SM prediction, can be accommodated. The ratio of the enhancements in the diphoton vs. the $WW$ and $ZZ$ channels cannot exceed 1.4. The $h\to b\bar{b}$ rate in associated production cannot exceed the SM rate by more than 50\%.}
\maketitle

\section{Introduction}
\label{sec:introducion}

Current LHC data have provided no evidence for beyond Standard Model (SM) physics, in particular, supersymmetry (SUSY). Instead, lower bounds were set on weak-scale SUSY, especially on the masses of the gluino and first two generation squarks~\cite{cmssusy, atlassusy}. It is not inconceivable that, with time, we will learn that the weak scale is tuned~\cite{ArkaniHamed:2004fb}. 

Nevertheless, the game is not over for naturalness~\cite{Papucci:2011wy}. Light stops and sbottoms have only been excluded up to about 200 GeV~\cite{Papucci:2011wy, Kats:2011qh, BRust:2011tb, Essig:2011qg, Aad:2011cw, Aad:2012}. Stronger limits exist but are model dependent, and it is still easy to find regions in the squark-neutralino-chargino parameter space where these limits cease to apply, with no need to invoke tuned degeneracies. On the theory side, rather than excluding naturalness, it is a consistent approach to take the experimental constraints as guidelines for model building of natural soft SUSY breaking. In particular, {\em flavored} SUSY breaking, with squarks of the first two generations much heavier than those of the third, gains motivation. Attempts to achieve this spectrum include~\cite{Dimopoulos:1995mi, Cohen:1996vb, Auzzi:2011eu, Csaki:2012fh, Craig:2012di, Craig:2012hc}. It is also conceivable that SUSY is natural and all squark generations are  light, but the collider signatures are altered so that the current searches are evaded. For example, the classic missing energy signature may not apply. Models in this spirit include $R$-parity violating SUSY~\cite{Csaki:2011ge, Graham:2012th}, stealth SUSY~\cite{Fan:2011yu, Fan:2012jf}, compressed SUSY~\cite{Alwall:2008ve, Alwall:2008va, LeCompte:2011cn, LeCompte:2011fh} and supersoft SUSY~\cite{Kribs:2012gx}.

In light of this debate between paradigms, the recent hint of a 125 GeV Higgs, reported by CMS~\cite{CMS, CMSphoton, CMSDiboson}, ATLAS~\cite{ATLAS, ATLASphoton, ATLASDiboson} and possibly also by the Tevatron experiments~\cite{TEVNPH:2012ab}, is exciting~\cite{Baer:2011ab, Heinemeyer:2011aa, Arbey:2011ab, Arbey:2011aa, Kadastik:2011aa, Moroi:2011aa, Buchmueller:2011ab, Cao:2011sn, Ellwanger:2011aa, Kang:2012tn, Olive:2012it, Ellis:2012aa, Baer:2012uy, Cao:2012fz, Jegerlehner:2012ju, Kang:2012ra, Curtin:2012aa, Cohen:2012zz, Christensen:2012ei, Boudjema:2012cq, Fowlie:2012im,  Gunion:2012zd, King:2012is, Rizzo:2012rf, Chang:2012gp, Vasquez:2012hn, Gupta:2012mi}.  
In the minimal supersymmetric SM (MSSM), a 125 GeV Higgs would imply fine-tuning at the few parts per thousand or so~\cite{Draper:2011aa,Hall:2011aa}. 
If SUSY is to be natural then it cannot be minimal, namely, new interactions beyond the MSSM must deform the Higgs sector.

Even if natural SUSY exists, given its evasive nature so far there is no guarantee that we shall see direct signatures of new physics any time soon. However, data have already begun to accumulate for one particular new set of measurements, namely, the production cross sections and decay rates of the Higgs boson. This data set is rich: a 125 GeV SM-like Higgs supports $\mathcal{O}(10)$ experimentally independent production and decay channels, including e.g. gluon fusion (GF) and vector boson fusion (VBF) production and decays such as $h\to\gamma\gamma,\,WW,\,ZZ,\,b\bar b,\,\tau\bar\tau$. This upcoming information provides promising means to probe indirectly the existence and details of natural SUSY. Given that the experimental uncertainties for each individual channel are large, the degree by which one can draw information from this new data set depends crucially on the degree of predictive power provided by the theory. This sets the scope for our current analysis.

In the MSSM, the third-generation sfermion, higgsino, gaugino, and Higgs sectors contain many free soft SUSY breaking parameters that can affect the Higgs couplings.
The question arise what definite predictions can be made if naturalness (together with experimental constraints) is used as guide.
The question gains depth given that a non-minimal Higgs sector would plausibly introduce additional free parameters in the scalar potential.

In this paper we attempt to partially address this question. Our working assumptions are:
\begin{enumerate}
\item We assume that an $m_h\approx$125 GeV Higgs exists, and is embedded in a natural (and therefore non-minimal) supersymmetric model;
\item While we do not commit to the MSSM quartic potential, we still assume that the weak scale Higgs sector is described by an approximately type-II two Higgs doublet model (2HDM);
\item We neglect beyond-MSSM physics in loops, assuming that the leading loop corrections to the Higgs-SM couplings involve MSSM fields only.
\end{enumerate}
None of these assumptions is necessarily true, but with them we obtain a predictive framework and show that it can be definitively tested against Higgs data. 

The outline of the paper is as follows. 
In general, we study the modifications to the Higgs couplings to SM particles, denoted by 
\beq\label{eq:r}
r_{i}\equiv\frac{g_{hii}}{g^{\rm SM}_{hii}}, 
\eeq
with $i = t, V, G, \gamma, b, \tau$ standing for top, massive vector gauge boson, gluon, photon, bottom and tau respectively\footnote{While this may not be strictly necessitated by data~\cite{Farina:2012ea}, we make the further assumption that $r_W=r_Z=r_V$.}. Our main task is to organize existing computations into concrete predictions, distinguishing relevant from irrelevant contributions in our framework. 

In Sec.~\ref{sec:loop} we consider MSSM loops including effects from stops, sbottoms, staus, gauginos, higgsinos and Higgs. Defining a measure of fine-tuning and demanding it not to exceed about 1:10, we argue that only stops and charginos can induce coupling modifications larger than $\sim5\%$. The most relevant constraints on the stop contribution to $r_G$ are electroweak precision tests and direct searches. Demanding less than $4\sigma$ tension in $(\Delta\rho/\rho)$, the  correction is limited to $r_G^{\tilde t}<1.3$. Rare $B$ decays constrain stop mixing and we incorporate $BR(B\to X_s \gamma)$ into our assessment of the model fine-tuning, finding that $r_G^{\tilde t}>0.85$ is preferred for naturalness. 
Very light charginos can vary the $h\gamma\gamma$ vertex in the range $0.7<r_\gamma^{\tilde\chi^\pm}<1.1$; the effect is limited by direct searches. 

In Sec.~\ref{sec:hmix} we consider Higgs mixing. We show that once one assumes that the weak-scale Higgs sector is an approximately type-II 2HDM, with natural flavor conservation broken only by $\tan\beta$-enhanced, but otherwise small, loop effects, then the analysis of Higgs observables involves only four free parameters. 
Because of the $\approx64\%$ branching fraction of a 125 GeV SM-like Higgs to $b\bar b$, the coupling $r_b$ is of key phenomenological importance. 
We estimate the possible variations in $r_b$, expected in concrete SUSY models that fall in our framework. Non-decoupling $D$-term models, obtained by integrating out additional gauge interactions, generically predict $r_b>1$. The precise result depends on the mass of the heavier Higgs doublet; for $m_H=350$ GeV, it translates into a $\sim30\%$ reduction in $BR(h\to \gamma\gamma)$, already in some tension with data. Non-decoupling $F$-term models, consisting of new chiral superfields with hypercharge zero, predict $r_b<1$.

In Sec.~\ref{sec:data} we give some natural SUSY predictions and compare them to  data. 
In particular, we show that (i) significant enhancement of the $h\to\gamma\gamma$ rate, up to a factor $\sim4$ times the SM value, is viable; (ii) the enhancement in $h\to\gamma\gamma$ cannot exceed the enhancement in $h\to ZZ,WW$ by more than 40\%, in some tension with LHC data; (iii) the enhancement in VBF production cannot exceed the enhancement in GF production by more than 50\%; (iv) the hint for $\sim100\%$ enhancement in $h\to b\bar b$ at the Tevatron is incompatible with natural SUSY.

We conclude in Sec.~\ref{sec:conc}. The appendix contains a discussion of a 125 GeV Higgs' implications for non-decoupling gauge extensions of the MSSM.

\section{Loop effects}
\label{sec:loop}
In this section we discuss the implications of light superpartners for Higgs production and decay channels at hadron colliders.  
We begin by summarizing the results.

The effects we consider include loop contributions from the charged Higgs, higgsinos, gauginos, stops, sbottoms, and staus. Of these, the only quantitatively relevant effects (potentially larger than $\sim5\%$) involve stops and charginos that affect the Higgs couplings to photons and gluons. 
The stop contribution to the Higgs-gluon coupling, $r_G^{\tilde t}$, is directly related and opposite in sign to the stop contribution to the Higgs photon coupling, $r_\gamma^{\tilde t}$; see Eq.~(\ref{eq:gamG}). The chargino contribution to $r_\gamma$ decouples for $\tan\beta\gsim5$, as the chargino-Higgs coupling scales linearly with $(1/\tan\beta)$.

\begin{figure}[!h]\begin{center}
\includegraphics[width=0.45\textwidth]{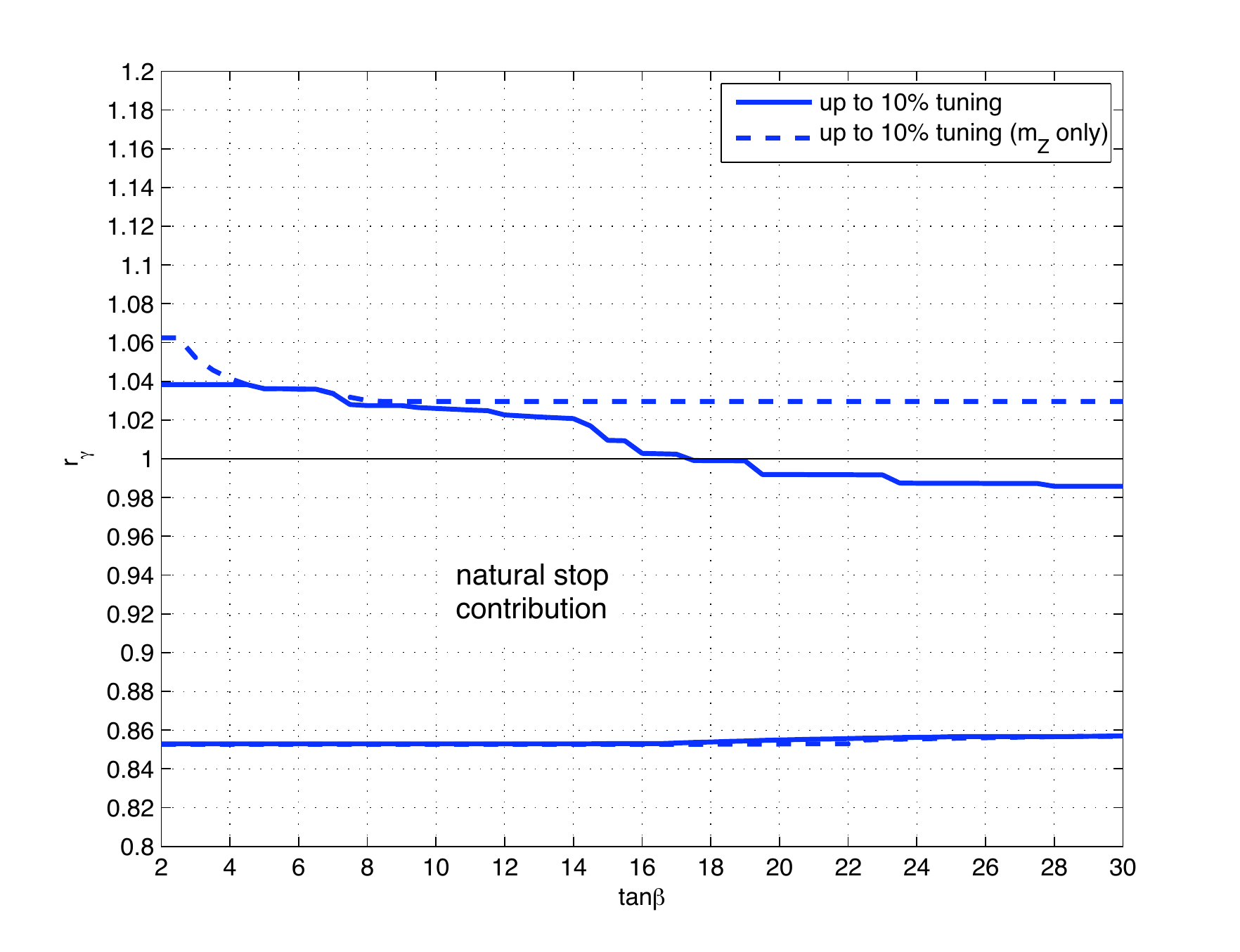}\quad
\includegraphics[width=0.45\textwidth]{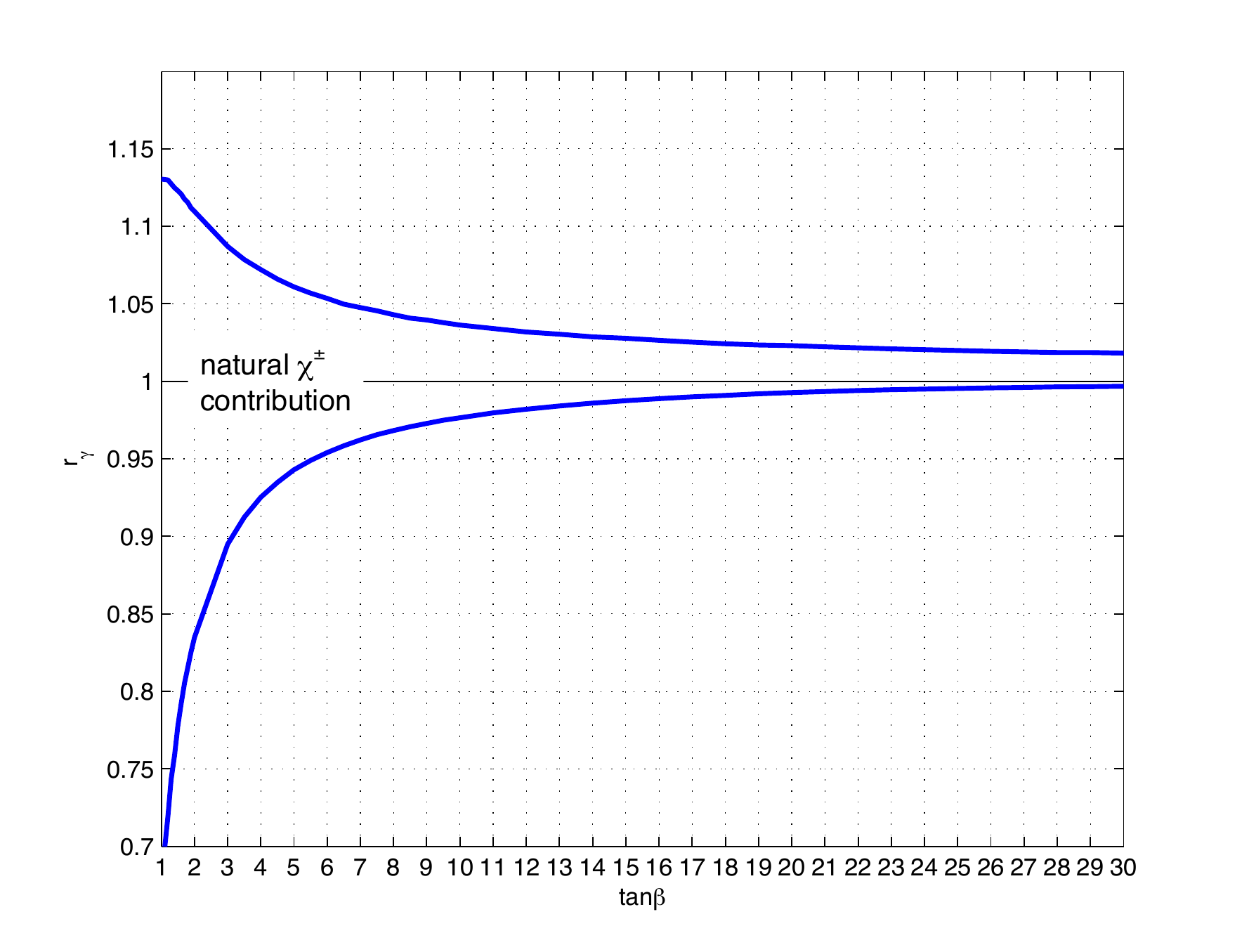}  
\end{center}
\caption{The range in $r_\gamma$ as a function of $\tan\beta$, compatible with fine-tuning no worse than 1:10 (details in Secs.~\ref{sec: stopsummary} and~\ref{ssec:chino}). Left: stop contribution. Right: chargino contribution. In the left panel, solid curves correspond to total fine-tuning, defined in Eq.~(\ref{eq:tot}), while dashed curves correspond to tuning with respect to the $Z$ boson mass alone, defined in Eq.~(\ref{eq:Dz}). Note that the stop contribution to $r_\gamma$ (left panel) is inversely related to $r_G$, see Eq.~(\ref{eq:gamG}).}
\label{fig:chino}
\end{figure}

\begin{figure}[!h]\begin{center}
\includegraphics[width=0.45\textwidth]{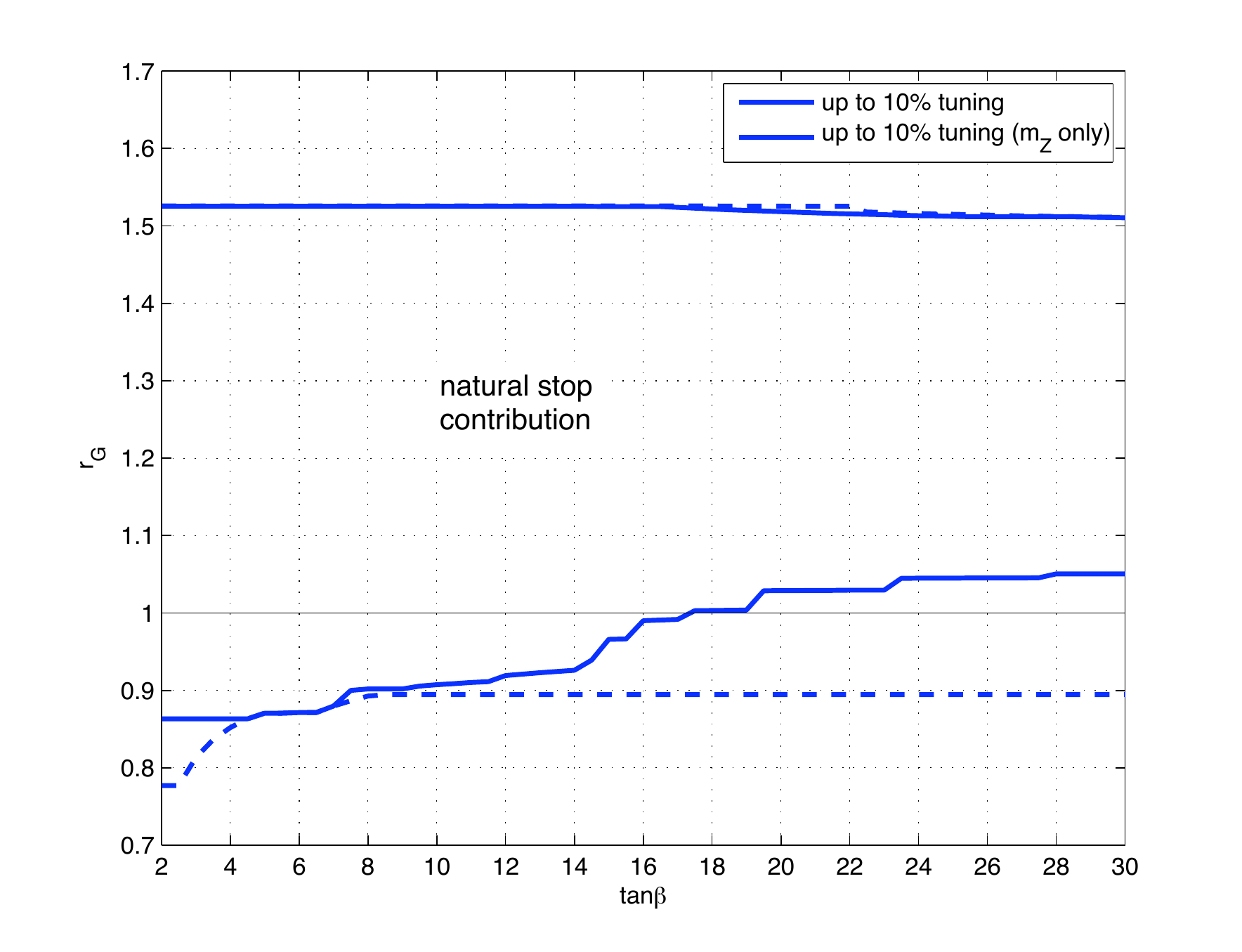}
\end{center}
\caption{Natural range in $r_G$ as a function of $\tan\beta$ (details in Sec.~\ref{sec: stopsummary}). Solid curves correspond to total fine-tuning, defined in Eq.~(\ref{eq:tot}), while dashed curves correspond to tuning with respect to the $Z$ boson mass alone, defined in Eq.~(\ref{eq:Dz}). Note that the stop contribution to $r_G$ is inversely related to $r_\gamma$, see Eq.~(\ref{eq:gamG}).}
\label{fig:FTrG2}
\end{figure}

The stop and chargino effects are constrained by direct searches for these particles and, to some extent, by naturalness. 
Imposing collider limits together with demanding fine tuning no worse than $1:10$, we check the possible size of these effects by varying the relevant theory parameters.  
The results are plotted in Fig.~\ref{fig:chino}, where the lines show the maximal and minimal values of $r_\gamma$ obtained within natural SUSY. The possible stop contribution to $r_G$ is plotted in Fig.~\ref{fig:FTrG2}.

The ranges of the different contributions to $r_G$ and $r_\gamma$ in natural SUSY are summarized in Table.~\ref{tab:range1}, limiting to $\tan\beta\geq2$. We also report upper limit estimates on the subdominant contribution of staus and sbottoms. In the following subsections we provide more details of the calculations. 
\begin{table}[h]
\begin{center}
\begin{tabular}{|c|c|c|}
\hline
& $r_G$ & $r_\gamma$ \\
\hline
${\tilde{t}}$ & $1^{+0.5}_{-0.14}$ & $1^{+0.04}_{-0.15}$ \\
\hline
${\tilde{\chi}^\pm}$ & $1$ & $1^{+0.1}_{-0.16}$ \\
\hline
$\tilde{b}$ & $1^{+0}_{-0.01}$ & $1^{+0.01}_{-0}$ \\
\hline
$\tilde{\tau}$ & $1$ & $1^{+0.03}_{-0}$ \\
\hline
\end{tabular}
\caption{Ranges of $r_G$ and $r_\gamma$ in natural SUSY from different  contributions, limiting to $\tan\beta\geq2$. The stop contributions to $r_G$ and $r_\gamma$ are anti-correlated, see Eq.~(\ref{eq:gamG}).}
\label{tab:range1}
\end{center}
\end{table}

\subsection{Stops}
\label{sec:stop}

\subsubsection{Higgs mass and fine-tuning}
\label{sec:stopmh}

The most widely studied radiative correction associated with stops is the increase in the Higgs mass. A well known, simple, analytic one-loop estimate that nevertheless includes the most relevant two-loop correction is given by~\cite{Drees:2004jm}
\beq\label{eq:mhst} m_h^2&\approx&m_Z^2c^2_{2\beta}+\frac{3G_F}{\sqrt{2}\pi^2}\left[m_t^4(Q_1)\log\left(\frac{M_s^2}{m_t^2}\right)+m_t^4(Q_2)\frac{X_t^2}{M_s^2}\left(1-\frac{X_t^2}{12M_s^2}\right)\right].\eeq
Here, $X_t=A_t-\mu/\tan\beta$, $M_s^2=m_{\tilde t_1}m_{\tilde t_2}$, $Q_1=\sqrt{m_tM_s}$, $Q_2=M_s$, and $m_t(Q)$ is the running top mass taken at scale $Q$,
\beq m_t(Q)=M_t\left(1-\frac{4\alpha_s(M_t)}{3\pi}\right)\left(\frac{\alpha_s(Q)}{\alpha_s(M_t)}\right)^{12/23}\eeq
with $M_t$ being the top pole mass, taken here as $172.5$ GeV.
Eq.~(\ref{eq:mhst}) slightly overestimates the Higgs mass, at a level of $\sim2$ GeV, compared to the numerical two-loop package FeynHiggs~\cite{Ellis:2007ka}. Here in part of the analysis we will use Eq.~(\ref{eq:mhst}) as a tight but robust upper bound to the Higgs mass in the MSSM, while in other parts we prefer the full FeynHiggs calculation, advising the reader accordingly.
 
The correction to the Higgs quartic coupling, leading to Eq.~(\ref{eq:mhst}), is accompanied by a shift to the Higgs self-energy that must be balanced by the bare mass in order not to destabilize the electroweak scale. Defining a fine-tuning measure, $\Delta_Z$, with respect to the $Z$ boson mass~\cite{Barbieri:1987fn}, we approximate the stop contribution to  $\Delta_Z$ by
\beq\label{eq:Dz}
\left(\Delta_Z^{-1}\right)_{\tilde t}=\left|\frac{2 \delta m_{H_u}^2}{m_Z^2}\right|,\;\;\;\quad \delta m_{H_u}^2|_{stop}=-\frac{3}{8\pi^2}y_t^2\left(m_{Q_3}^2+m_{u3}^2+A_t^2\right)\log\left(\frac{\Lambda}{{\rm TeV}}\right).
\eeq
In what follows we assume conservatively a low SUSY breaking mediation scale $\Lambda$ = 20 TeV.

Eqs.~(\ref{eq:mhst}) and (\ref{eq:Dz}) imply that a Higgs mass of 125 GeV would be highly fine-tuned in the MSSM. Unless we let go of natural SUSY, some physics beyond the MSSM must deform the scalar Higgs sector. Under this assumption, Eq.~(\ref{eq:mhst}) does not directly constrain the stop sector. However, Eq.~(\ref{eq:Dz}) continues to provide a reasonable guide to the fine-tuning associated with stops, with typically only mild modifications that we discuss in Secs.~\ref{sec:Dterm} and~\ref{sec:Fterm}, in the context of concrete examples. 

\subsubsection{Higgs couplings to gluons and photons}
\label{sec:stopcou}
The Higgs low energy theorem tells us that in the presence of heavy colored multiplets whose masses depend on the Higgs vacuum expectation value (VEV), e.g., the top-stop multiplet, the leading log Higgs-gluon-gluon coupling at the one-loop level is given by~\cite{Shifman:1979eb,Dermisek:2007fi}
\beq
{\cal L}_{hgg} = \frac{\alpha_s}{12\pi}\frac{h}{v}\left(2\sum_{F}t_{F}\frac{\partial \log {\rm det} \,m_{F}(v)}{\partial \log v}+\frac{1}{2}\sum_{S}t_{S}\frac{\partial \log {\rm det} \, m_{S}(v)}{\partial \log v}\right) G_{\mu\nu}^a G^{a \mu\nu}, \label{eq:general_hgg} 
\eeq
where $F$ and $S$ denote respectively colored fermion and scalar with Dynkin index $t_{X}$ ($=1/2$ for the fundamental representation) and mass matrices $m(v)$. Applying Eq.~(\ref{eq:general_hgg}) to the top-stop multiplet, we have~\cite{Arvanitaki:2011ck}
\beq\label{eq:rG}
r_G^{\tilde t}-1
\approx \frac{1}{4} \left(\frac{m_t^2}{m_{\tilde{t}_1}^2}+\frac{m_t^2}{m_{\tilde{t}_2}^2}-\frac{m_t^2X_t^2}{m_{\tilde{t}_1}^2m_{\tilde{t}_2}^2}\right), \quad {\rm stop \, contribution,}
\eeq
where we neglect $D$-terms. We factored out an overall correction factor $r_t$, defined via Eq.~(\ref{eq:r}), that comes about by Higgs mixing. The total $hGG$ vertex correction reads
\beq\label{rGtot} r_G=r_t\,r_G^{\tilde t}.\eeq
%
Eq.~(\ref{eq:rG}) compares well with results from FeynHiggs; nevertheless, in numerical computations we include the $D$-term contribution. Concerning the leading log approximation, this can be checked by comparing the full fermion and scalar loop function ratio evaluated at $m_t$ and $m_{\tilde t}$,\\
 $\left[\mathcal{F}_0\left(m_h^2/4m_{\tilde t}^2\right)/\mathcal{F}_{1/2}\left(m_h^2/4m_t^2\right)\right]$, to the asymptotic value (1/4) of this ratio at $m_h\to0$. Varying $m_{\tilde t}$ between 150-1000 GeV, we find that the leading log approximation is good to about 6\%. 

For reasonably light stops, Eq.~(\ref{eq:rG}) leads to a substantial effect, e.g. with $m_{\tilde{t}_1} = m_{\tilde{t}_2} = 250$ GeV, $r_G = 1.24$, implying 53\% increase in GF production. As long as stop mixing is small, the $hGG$ coupling is enhanced compared to the SM and consequently the GF rate is enhanced. As discussed in~\cite{Dermisek:2007fi}, large $X_t$ could in principle reduce the $hGG$ coupling. However, naturalness, together with the direct bound $m_{\tilde t_1}>100$ GeV, limit this possibility: large $X_t$ adds to the weak-scale fine tuning both directly, through Eq.~(\ref{eq:Dz}), and indirectly because it requires a larger diagonal soft mass to start with. In the next section we exhibit further constraints on such large $X_t$ that arise from rare $B$ decays at large $\tan\beta$. 

There is an inverse correlation between the top/stop contributions to the Higgs effective coupling to photons and to gluons, the negative sign coming because of the dominant $W$ diagram that contributes to $h\gamma\gamma$ with opposite sign from the matter loops. To see this, let us denote the $W$ and top loop contributions to the $h\gamma\gamma$ amplitude by $\mathcal{A}^\gamma_W$ and $\mathcal{A}^\gamma_t$, respectively, and the stop contribution by $\mathcal{A}^\gamma_{\tilde t}$. Let us further define the $hGG$ top and stop-induced amplitudes by $\mathcal{A}^G_t$ and $\mathcal{A}^G_{\tilde t}$, and note that 
\beq\frac{\mathcal{A}^\gamma_{\tilde t}}{\mathcal{A}^\gamma_t}=\frac{\mathcal{A}^G_{\tilde t}}{\mathcal{A}^G_t}=r_G^{\tilde t}-1,\eeq
to leading order in $\alpha_s$.
This gives
\beq\label{eq:gamG} r_\gamma=\frac{\mathcal{A}^\gamma_W+\mathcal{A}^\gamma_t+\mathcal{A}^\gamma_{\tilde t}}{\left(\mathcal{A}^\gamma_W+\mathcal{A}^\gamma_t\right)^{\rm SM}}
\approx1.28r_V-0.28r_G,\quad W,{\rm\,top,\,and\,stop\;contributions,}\eeq
using $\mathcal{A}^\gamma_W\approx8.33$ and $\mathcal{A}^\gamma_t\approx-1.84$ in the SM, valid for $m_h=125$ GeV.

Eqs.~(\ref{rGtot}) and~(\ref{eq:gamG}) do not include loop contributions of additional particles, notably charginos and bottom and tau fermions and scalars. The bottom and tau fermion contributions remain below about five percent of the top even for $r_{b,\tau}\sim10$. The chargino, sbottom and stau contributions can in principle become relevant in some corners of the MSSM parameter space, resulting with some loss of predictivity by disturbing the $r_\gamma-r_G$ correlation of Eq.~(\ref{eq:gamG}). Below we examine these terms in more detail, concluding that in natural SUSY, the sbottom and stau contributions can be neglected while charginos may lead to marginally observable effects.

\subsubsection{Large stop mixing vs. fine-tuning in $BR(B\to X_s\gamma)$}
\label{sec: btosgamma}
Light, mixed stops are constrained by rare $B$ decays. 
The branching fraction for the rare decay $B\to X_s\gamma$ has been measured experimentally to a precision of better than ten percent~\cite{Asner:2010qj},
\beq\label{eq:expbsg}BR(B\to X_s\gamma)^{\rm exp}&=&(3.52\pm0.25)\times10^{-4}.\eeq
The theoretical SM NNLO calculation has reached a similar accuracy~\cite{Benzke:2010js}\footnote{Ref.~\cite{Misiak:2006zs} found the theoretical result $BR(B\to X_s\gamma)^{\rm SM}=(3.15\pm0.25)\times10^{-4}$.},
\beq\label{eq:smrbsg}BR(B\to X_s\gamma)^{\rm SM}&=&(2.98\pm0.26)\times10^{-4}.\eeq
The theoretical NNLO SM prediction is fully determined by observable quantities, namely the masses of the top quark and $W$ boson and gauge couplings. Therefore, the agreement (within $\sim1.5\sigma$) between Eqs.~(\ref{eq:expbsg}) and~(\ref{eq:smrbsg}) allows us to define an observable quantity, $\OO_{bs\gamma}$, that we can compare against models of new physics,
\beq\label{eq:Obsg}\OO_{bs\gamma}&=&\frac{BR(B\to X_s\gamma)^{\rm exp}}{BR(B\to X_s\gamma)^{\rm SM}}-1=0.18\pm0.13.\eeq

Eq.~(\ref{eq:Obsg}) means that new physics is now only allowed to contribute to $B\to X_s\gamma$ at about thirty percent of the SM contribution. Because the SM contribution begins at one-loop, new physics models such as SUSY can easily produce larger contributions. Recalling the possibility of accidental cancellations, typical SUSY Higgs analyses in the literature either ignore $B\to X_s\gamma$ or  focus on parameter regions where cancellations occur. Here, given our interest in natural models, we will use Eq.~(\ref{eq:Obsg}) to estimate the level of fine-tuning involved in the latter approach~\cite{Perelstein:2007nx}.

Given a model input parameter $P$ (e.g., $A_t$) that contributes to $BR(B\to X_s\gamma)$, we assess the degree of fine-tuning $\Delta$ associated with it in a similar way to the fine-tuning measure commonly associated with the $Z$ boson mass. The only slight modification we apply here is to account for the uncertainty in the experimental determination of $\OO_{bs\gamma}$:
\beq\label{eq:Dbsg}\Delta_{\OO_{bs\gamma}}^{-1}&=&\left|\frac{\OO_{bs\gamma}}{\sigma_{\OO_{bs\gamma}}}\frac{\partial\log\OO_{bs\gamma}}{\partial \log P}\right|=\left|\frac{P}{0.3}\frac{\partial\OO_{bs\gamma}}{\partial P}\right|,\eeq
where we chose to combine linearly the absolute values of the central value and of the uncertainty on the right hand side of Eq.~(\ref{eq:Obsg}), setting $\sigma_{\OO_{bs\gamma}}=0.3$. A total fine tuning is defined as
\beq\label{eq:tot}
\Delta_{tot}^{-1}=\sqrt{\Delta_{Z}^{-2}+\Delta_{\OO_{bs\gamma}}^{-2}}.
\eeq

Consider now minimally flavor violating new physics contributions to the Wilson coefficients $C_{7,8}$ of the electromagnetic and chromomagnetic dipole operators $\OO_{7,8}$,
\beq
\OO_7=\frac{e}{16\pi^2}m_b \left(\bar{s}_L\sigma_{\mu\nu}b_R\right)F^{\mu\nu}, \quad
\OO_8=\frac{g}{16\pi^2}m_b \left(\bar{s}_L\sigma_{\mu\nu}T^ab_R\right)G^{a\mu\nu}.
\eeq
Taking $C_{7,8}$ to be input at the top mass scale, the contribution to $\OO_{bs\gamma}$ can be approximated by~\cite{Lunghi:2006hc} 
\beq\label{eq:wc}\OO_{bs\gamma}&=&a_{77}|C_7|^2+a_{88}|C_8|^2+\Re\left\{a_7C_7+a_8C_8+a_{78}C_7C_8^*\right\},\eeq
with $a_7=-2.41+0.21i,\,a_8=-0.75-0.19i,\,a_{77}=1.59,\,a_{88}=0.26,\,a_{78}=0.82-0.30i$. 
The MSSM with light mixed stops and light higgsinos makes a $\tan\beta$-enhanced contribution to the Wilson coefficients, of the form
\beq\label{eq:C78} C_{7,8}&\approx&\frac{m_t^2A_t\mu}{m_{\tilde t}^4}\, \FF_{7,8}\left(\frac{m^2_{\tilde t_1}}{|\mu|^2},\frac{m^2_{\tilde t_2}}{|\mu|^2}\right)\,\tan\beta,\eeq
where $\FF_{7,8}(x,y)$ are loop functions that take $\OO(1)$ values for $x\sim y=\OO(1)$. Using Eqs.~(\ref{eq:C78}),~(\ref{eq:wc}) and~(\ref{eq:Dbsg}) we see that if we wish to avoid accidental cancellations to a level of one part per ten, then we must have $(A_t\mu\tan\beta/m_{\tilde t}^2)<$ few. 
In our analysis we compute the stop contribution to $\OO_{bs\gamma}$, and exclude a parameter space from being part of natural SUSY based on a tuning criterion, such as $\Delta_{tot}<10\%$, rather than on the precise deviation of $\OO_{bs\gamma}$ from its central value.

\subsubsection{Summary of stop effects}
\label{sec: stopsummary}
In Figs.~\ref{fig:chino} (left) and~\ref{fig:FTrG2} we study the range of values for $r_\gamma^{\tilde t}$ and $r_G^{\tilde t}$ that can be naturally expected due to light stops. Here, we plot the extremal (upper and lower) values of $r_\gamma^{\tilde t}$ and $r_G^{\tilde t}$ that can be achieved by varying $A_t,\mu,m_Q,m_U$ for different values of $\tan\beta$, while imposing $\Delta_{tot}>10\%$, or tuning no worse than one part per ten. We also, very conservatively, demand that $m_{\tilde t_1}>100$ GeV. 
The extra fine-tuning incurred by large $A_t$, as seen by Eqs.~(\ref{eq:Dz}) and~(\ref{eq:C78}), suggests that stop mixing should not be large, and more so for larger $\tan\beta$. As a result, the lower bound on $r_G^{\tilde t}$ (upper bound on $r_\gamma^{\tilde t}$) is set by naturalness.  

We learn that $r_G^{\tilde t}<0.85$ does not arise within natural SUSY, but a rather significant increase of $r_G^{\tilde t}\sim1.4$ is obtained with light, unmixed stops. 
The upper bound $r_G^{\tilde t}<1.5$ obtains when stops are un-mixed and as light as we let them, close to the top mass. Ignoring the possibility of additional SUSY contributions, this ultra-light region is in strong tension with electroweak precision tests (EWPTs). For example, setting $X_t=0,\,m_Q=m_U=100$ GeV ($m_{\tilde t_{1,2}}\approx200$ GeV) and neglecting sbottom mixing we find $(\Delta\rho/\rho)\approx3.5\times10^{-3}$, more than $8\sigma$ deviation. However, it is not inconceivable that contributions from the Higgs and gaugino sectors could ameliorate this tension. While such cancellation would certainly be fine-tuned, the level of tuning does not quite make it to the 1:10 level except at this very low mass end. Instead of complicating the analysis by folding in EWPTs into our naturalness criterion, we simply report that if we impose that the stop-sbottom contribution to $(\Delta\rho/\rho)$ remains within the $4\sigma$ range, we obtain stronger limits on the Higgs-gluon and Higgs-photon vertices, $0.9<r_G^{\tilde t}<1.3$ and $0.9<r_\gamma^{\tilde t}<1.03$. 

\begin{figure}[!h]
\begin{center}
\includegraphics[width=0.45\textwidth]{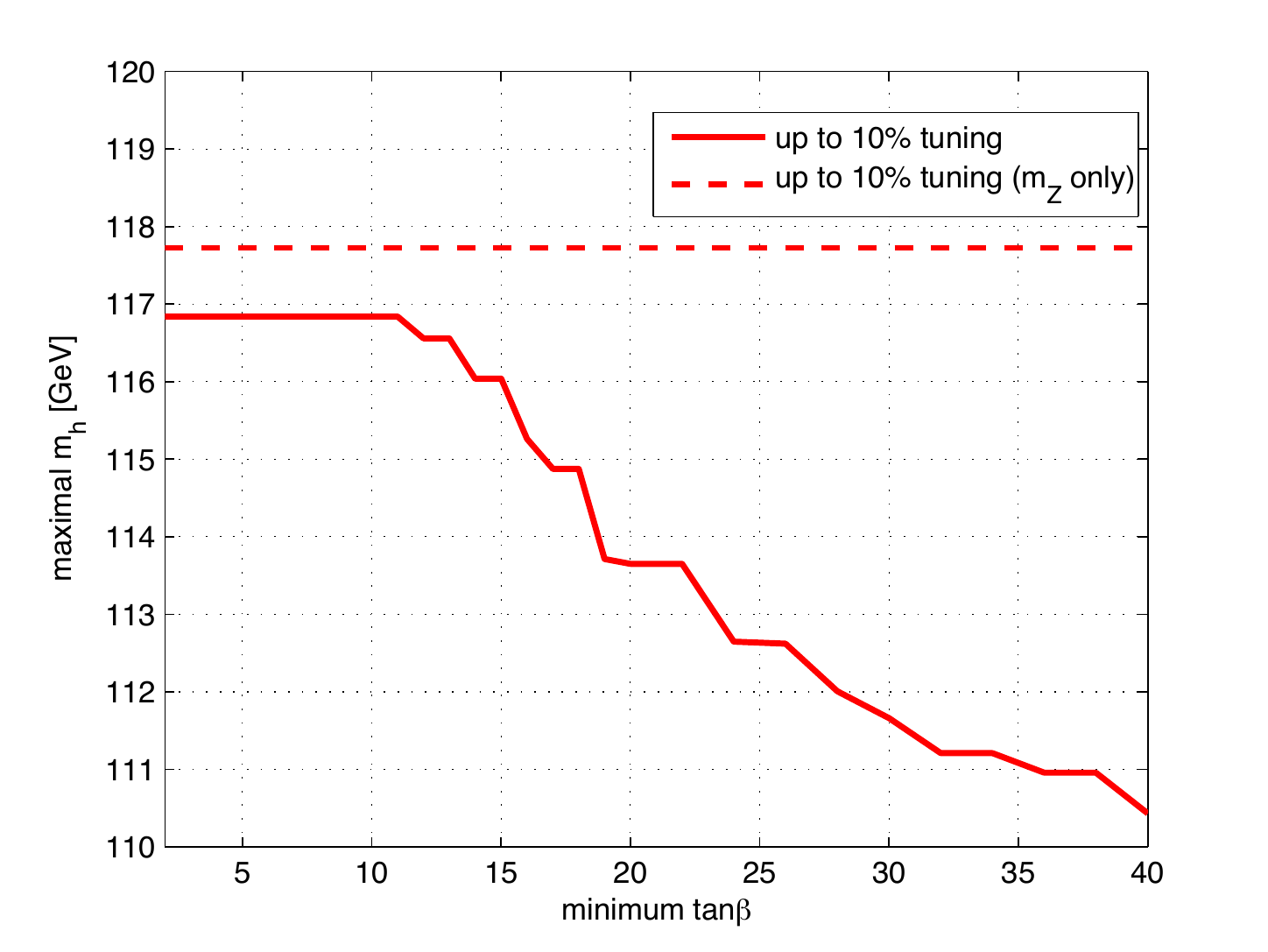}  
\end{center}
\caption{Upper limit on $m_h$ in the MSSM, as function of $\tan\beta$. The solid line corresponds to scenarios with up to 10\% total fine-tuning (defined in Eq.~(\ref{eq:tot})), while the dashed line corresponds to $Z$ boson mass tuning alone (Eq.~(\ref{eq:Dz})).}
\label{fig:stopmh}
\end{figure}

\begin{figure}[!h]
\begin{center}
\includegraphics[width=0.45\textwidth]{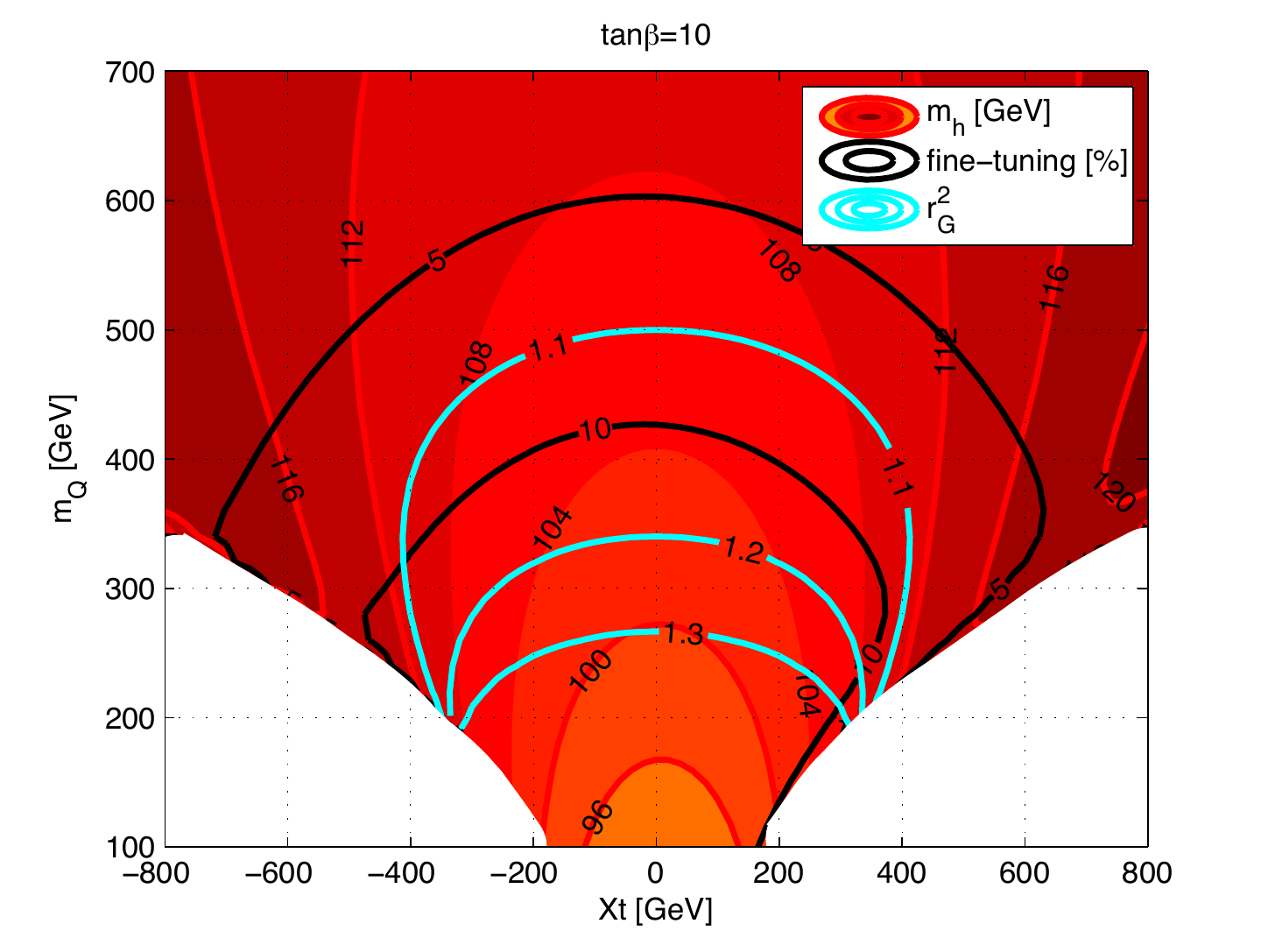}\quad
\includegraphics[width=0.45\textwidth]{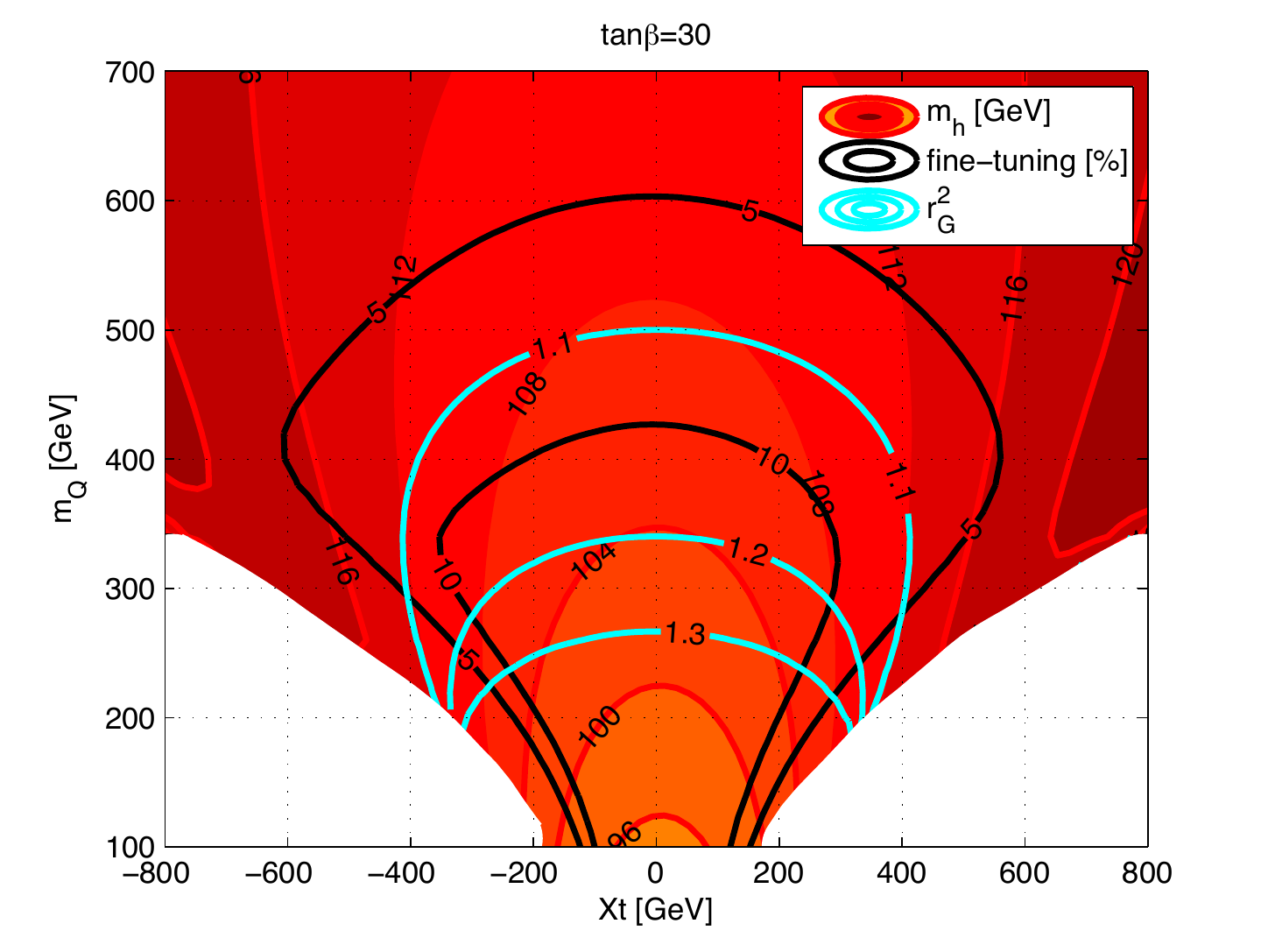}\end{center}
\caption{Contours of the Higgs mass, the total fine tuning and $r_G^2$ in the $(m_{Q_3}, X_t)$ plane. We set $\mu = 150$ GeV, $m_{Q_3} = m_{u_3}$ and $X_t=A_t-\mu/\tan\beta$.}
\label{fig:stop}
\end{figure}

Finally, while our philosophy in this work is that additional physics beyond the MSSM must affect the Higgs sector to account for the Higgs mass, it is nevertheless interesting to exhibit the implications of naturalness for the stop contribution to $m_h$. In Fig.~\ref{fig:stopmh} we plot the maximal value of $m_h$ obtained by varying the SUSY parameters as above. We use the analytical estimate for $m_h$ and so this plot provides an upper bound. The plot shows how tuning for $B\to X_s\gamma$ makes a large stop contribution to $m_h$ less plausible for large $\tan\beta$.
Fig.~\ref{fig:stop} shows contours of the MSSM Higgs mass (computed now using FeynHiggs), the total fine tuning defined in Eq.~(\ref{eq:tot}) and $\left(r_G^{\tilde t}\right)^2$ in Eq.~(\ref{eq:rG}), for two values of $\tan\beta=10,30$. At large $\tan\beta$, tuning for $B\to X_s\gamma$ disfavors the high mixing region, causing the maximal value of $m_h$ to drop substantially. 


\subsection{Charginos}\label{ssec:chino}

Naturalness dictates that at least one chargino must be light, $m_{{\tilde{\chi}}^\pm}\lsim200$ GeV. Hence, the chargino contribution to $r_\gamma$ may be expected to become relevant~\cite{Gunion:1988mf, Djouadi:1996pb, Diaz:2004qt}. What limits the effect to be modest is the direct bound, that we take to be $m_{{\tilde{\chi}}^\pm}>94\,{\rm GeV}$~\cite{Nakamura:2010zzi}. 
Imposing this bound, we compute the chargino contribution to $r_\gamma$, varying the relevant parameters in the range $-300<\mu/$GeV$<300$, $0<M_2/$GeV$<1000$, $1<\tan\beta<40$. The result is shown in the right panel of Fig.~\ref{fig:chino}. We conclude that:
\begin{itemize}
\item A sizable effect is possible only for low $\tan\beta<3$, mostly limited to a reduction in $r_\gamma$. The effect is largest for $\tan\beta=1$, where we find $0.7< r^{\tilde\chi^\pm}_\gamma<1.13$. If we restrict to $\tan\beta\geq2$, we have $0.8< r^{\tilde\chi^\pm}_\gamma<1.1$. The sign of $(r_\gamma^{\tilde\chi^\pm}-1)$ depends upon the sign of $\left(\mu M_2\right)$. 
\item Restricting to $\tan\beta>3$ ($\tan\beta>5$) diminishes the effect, as the chargino-Higgs coupling $\propto\sin2\beta$ is reduced. Here we find $|r^{\tilde\chi^\pm}_\gamma-1|<10\%(6\%)$, where saturating the upper limit requires two very light charginos with $m_{{\tilde{\chi}}^\pm_1}\sim m_{{\tilde{\chi}}^\pm_2}\sim100$ GeV.
\end{itemize}

The upper bound of $\sim+10\%$ for the chargino enhancement to the Higgs-photon coupling can be understood as follows. Taking $\tan\beta=1$ one obtains, at leading log,
\beq r_\gamma^{\tilde\chi^\pm}&\approx&\frac{-6.49-\frac{8}{3}\frac{m_W^2}{M_2\mu-m_W^2}}{-6.49}\leq1+\frac{8}{3\times6.49}\frac{m_W^2}{m^2_{\tilde\chi^\pm_1}}\left(1+\frac{2m_W}{m_{\tilde\chi^\pm_1}}\right)^{-1},\eeq
where we took $M_2\mu>m_W^2$ in order to obtain positive interference with the $W$-dominated SM amplitude $\mathcal{A}_{SM}\approx-6.49$. Imposing the bound $m_{\tilde\chi^\pm_1}>94$~GeV, we obtain $r_\gamma^{\tilde\chi^\pm}-1\lsim10\%$, in good agreement with the full one-loop computation.

The bottom line is that a chargino contribution to the $h\gamma\gamma$ vertex, with a sign that is theoretically unconstrained, can disturb the correlation between $r_\gamma$ and $r_G$ that we have found, accounting for the stop contribution alone. As a result we will be forced to assess $r_G$ and $r_\gamma$ independently when we come to predict Higgs observables. Nevertheless, it will still be useful to describe the modified $h\gamma\gamma$ and $hGG$ in terms of $r^{\tilde t}_G$ and $r_\gamma^{\tilde\chi^\pm}$. This separation becomes practical for $\tan\beta\gsim3$, when the chargino correction decouples, with ramifications to Higgs mixing effects.

\subsection{Charged Higgs}\label{ssec:ch}
A charged Higgs loop diagram contributes to $r_\gamma$, with~\cite{Blum:2012kn}
\beq r_\gamma\approx1-0.007\left(\frac{\lambda_{hH^+H^-}}{\lambda_{hH^+H^-}^{MSSM}}\right)\left(\frac{m_{H^\pm}}{250\,\rm GeV}\right)^{-2},\eeq
where $\lambda_{hH^+H^-}^{MSSM}=(g^2-g'^2)/4\approx0.07$ is the MSSM coupling.
The contribution is negligible unless the charged Higgs is very light, in strong tension with $B\to X_s\gamma$~\cite{Misiak:2006zs}\footnote{The tension with $B\to X_s\gamma$ may be ameliorated if the charged state is taken from an additional inert Higgs multiplet.}, or the coupling $\lambda_{hH^+H^-}$ receives very large corrections from an extended Higgs sector. We note that in $F$-term models where singlet chiral superfields are added, to be discussed in more detail shortly, a numerically large correction to some Higgs quartic couplings is conceivable; however, the coupling $\lambda_{hH^+H^-}$ remains unaffected.

\subsection{Staus and sbottoms}\label{ssec:stau}
The possibility that light scalar $\tau$ could boost the $h\gamma\gamma$ coupling was entertained in~\cite{Carena:2011aa, Carena:2012gp}. This possibility is, however, outside of the scope for natural SUSY. Naturalness limits this effect as follows. The stau mass eigenvalues are
\beq m_{\tilde \tau_{1,2}}^2\cong\frac{m_{\tilde L}^2+m_{\tilde e}^2+\frac{m_Z^2c_{2\beta}}{2}}{2}\pm\sqrt{\frac{\left(m_{\tilde L}^2-m_{\tilde e}^2+2m_Z^2s_W^2c_{2\beta}\right)^2}{4}+m_\tau^2X_{\tilde\tau}^2},\eeq
with $X_{\tilde\tau}=A_{\tilde\tau}-\mu\tan\beta$.
Let us neglect $D$-terms in the following discussion, for clarity, including them numerically later to verify our conclusions. The leading log contribution to $r_\gamma$ can then be estimated as
\beq r_\gamma -1 \approx-\frac{1}{6\mathcal{A}_{SM}}\frac{\partial\log \left(m_{\tilde \tau_{1}}^2m_{\tilde \tau_{2}}^2\right)}{\partial\log v}=1+\frac{r_\tau}{3\mathcal{A}_{SM}}\frac{m_\tau^2X_{\tilde\tau}^2}{m_{\tilde \tau_{1}}^2m_{\tilde \tau_{2}}^2}, \quad {\rm stau \, contribution},\eeq
where $\mathcal{A}_{SM}\approx6.49$ is the SM ($W$ and $t$ loop) amplitude, evaluated at $m_h=125$ GeV. Using the fact that $m_{\tilde \tau_{2}}^2\geq m_{\tilde \tau_{1}}^2+2m_\tau\left|X_{\tilde\tau}\right|$, we obtain
\beq 0<r_\gamma-1\lsim4r_\tau\times10^{-4}\left(\frac{\left|X_{\tilde\tau}\right|}{100\,{\rm GeV}}\right)\left(\frac{m_{\tilde \tau_{1}}}{100\,\rm GeV}\right)^{-2}, \quad {\rm stau \, contribution}. \eeq
Imposing conservatively $m_{\tilde\tau_1}>100$ GeV and demanding fine-tuning no worse than $5\%$ ($|\mu|\lsim300$ GeV), we find that in natural SUSY the effect is not larger than 3\% for $\tan\beta<50$ and $r_\tau\sim1$. 

The sbottom contribution to $r_\gamma$ is estimated to be even smaller because of the smaller electric charge, 
\beq 0<r_\gamma-1\lsim r_b\times10^{-4}\left(\frac{\left|X_{\tilde{b}}\right|}{100\,{\rm GeV}}\right)\left(\frac{m_{\tilde {b}_{1}}}{100\,\rm GeV}\right)^{-2}, \quad {\rm sbottom \, contribution}. \eeq
The contribution to the $hGG$ vertex is given by
\beq
r_G-1\approx-\frac{(r_b/r_t)}{4}\frac{m_b^2X_b^2}{m_{\tilde{b}_1}^2m_{\tilde{b}_2}^2}, \quad {\rm sbottom \, contribution,}
\label{eq:hggsb}
\eeq
that we can bound by
\beq 0>r_G-1\gsim-0.01(r_b/r_t)\left(\frac{\left|X_{\tilde{b}}\right|}{100\,{\rm GeV}}\right)\left(\frac{m_{\tilde {b}_{1}}}{100\,\rm GeV}\right)^{-2}, \quad {\rm sbottom \, contribution}. \eeq

To make the stau or sbottom contributions to $r_\gamma$ or $r_G$ larger than a few percent, large values of $r_\tau,r_b$, arising from Higgs mixing, would be required. As we show in the next sections, (i) one expects $r_b\sim r_\tau$; (ii) large $r_b$ would increase the total Higgs width, implying a suppression to the $h\to\gamma\gamma$ rate that would more than compensate for the presumed loop correction and that is not seen in the data; and (iii) $r_b\gsim3$ does not arise in the MSSM or any extension we are aware of for lifting the Higgs mass. We conclude that the sbottom and stau contributions to $r_\gamma$ and $r_G$ can be safely neglected in assessing the predictions of natural SUSY to a few percent accuracy.

\section{Higgs mixing}\label{sec:hmix}

Including quantum corrections, the natural Higgs mass prediction in the MSSM does not exceed 100-110 GeV (see Fig.~\ref{fig:stop}). A Higgs at 125 GeV then implies that quartic couplings in the scalar potential receive corrections at the $\sim50\%$ level from new physics beyond the MSSM. These corrections can modify the Higgs-fermion and Higgs-vector couplings from their MSSM values at a similar level, in a model dependent manner, through Higgs mixing. 
In this section we derive the Higgs coupling modifications due to Higgs mixing. We do not commit ourselves to the MSSM structure, making instead the more general assumption of a 2HDM framework. We then discuss the implications for specific model examples. 

\subsection{Two Higgs Doublet Model}
\label{sec:2hdm}
We do not consider here the possibility of additional light fields mixing with $H_{u,d}$, as could occur e.g. if the MSSM is augmented with a light singlet. In addition, we assume that non-renormalizable interactions in the effective 2HDM can be neglected. 
Then above the weak scale but below (at least most of) the superpartners, the supersymmetric Higgs sector is described by an approximately type-II 2HDM~\cite{Gunion:2002zf,BRanco:2011iw},
\beq\label{eq:V2hdm}-\mathcal{L}&=&H_1^\dag\mathcal{D}^2H_1+H_2^\dag\mathcal{D}^2H_2+m_1^2|H_1|^2+m_2^2|H_2|^2\no\\
&+&\frac{\lambda_1}{2}|H_1|^4+\frac{\lambda_2}{2}|H_2|^4+\lambda_3|H_1|^2|H_2|^2+\lambda_4|H_1\sigma_2 H_2|^2\no\\
&+&\Big\{\frac{\lambda_5}{2}(H_1^\dag H_2)^2+(H_1^\dag H_2)\left(m_{12}^2+\lambda_6|H_1|^2+\lambda_7|H_2|^2\right)\no\\
&+&Y_tH_2\epsilon \bar t_RQ_{L3}+\left(Y_bH_1^\dag-Y_b\Delta_bH_2^\dag\right)\bar b_RQ_{L3}+\left(Y_\tau H_1^\dag-Y_\tau \Delta_\tau H_2^\dag\right) \bar \tau_RL_3+h.c.\Big\}\,,\eeq
where $H_{1,2}\sim(1,2)_{+1/2}$ and we identify $H_2=H_u,\,H_1=i\sigma_2H_d^*$. 
In Eq.~(\ref{eq:V2hdm}) we include ``wrong" Higgs couplings for bottom and tau fermions with coefficients $\Delta_{b,\tau}$, that come about from integrating out third generation squarks, higgsinos and gauginos at one-loop. We omit first and second generation fermions.

Traditionally, the analysis of Higgs couplings is achieved by diagonalizing the Higgs mass matrix and expressing the couplings in terms of the rotation angle $\alpha$, connecting the interaction basis of Eq.~(\ref{eq:V2hdm}) to the mass basis, and the ratio~\cite{Gunion:2002zf}
\beq\tan\beta=\frac{\langle H_u\rangle}{\langle H_d\rangle}.\eeq
Omitting for the moment $\Delta_{b,\tau}$, we have $r_b=r_\tau$, and 
\beq\label{eq:rs2hdm}
r_b=\frac{vg_{hb\bar b}}{m_b}=-\frac{\sin\alpha}{\cos\beta},\;\;\;r_t=\frac{vg_{ht\bar t}}{m_t}=\frac{\cos\alpha}{\sin\beta},\;\;\;r_V=\frac{vg_{hVV}}{2m_V^2}=\sin\left(\beta-\alpha\right),
\eeq
implying the inequalities
\beq r_b^2\leq\tan^2\beta+1,\;\;\;r_t^2\leq\frac{1}{\tan^2\beta}+1,\;\;\;r_V^2\leq1.\eeq
 
We are free to choose two independent parameters to describe $r_b,\,r_t$ and $r_V$. We choose these parameters to be $\tan\beta$ and $r_b$. With this choice we write
\beq\label{eq:parr} r_t=\sqrt{1-\frac{r_b^2-1}{\tan^2\beta}},\;\;\;\;r_V=\frac{\tan\beta}{1+\tan^2\beta}\left(\frac{r_b}{\tan\beta}+\sqrt{1+\tan^2\beta-r_b^2}\right),\eeq
valid for all $\tan\beta$. 
We assumed that $r_t\geq0$, taking the positive root. 

We now comment on the validity of neglecting $\Delta_{b,\tau}$ in Eq.~(\ref{eq:parr}). As it turns out, $\Delta_{b,\tau}$, that enter the Higgs couplings $\tan\beta$-enhanced, only become quantitatively relevant at large $\tan\beta$. However, if $(r_b/\tan\beta)^2\ll1$, then deviations in $r_t,\,r_V$ are suppressed compared to deviations in $r_b$. As we discuss below in more detail, this results in the fact that whenever the values of $r_t$ or $r_V$ are non-negligible phenomenologically, then Eq.~(\ref{eq:parr}) applies to high accuracy even when finite $\Delta_{b,\tau}$ are introduced. This conclusion is useful: it means that for arbitrary new physics deformations of the MSSM Higgs potential -- just as long as the basic 2HDM structure is maintained  -- only two variables, $r_b$ and $\tan \beta$, are required to describe Higgs mixing effects on the lighter Higgs effective couplings. Note, finally, that similar diagrams to those that produce $\Delta_{b,\tau}$, also produce finite $\lambda_{6,7}$. As the latter couplings are vanishing in the MSSM, this can lead to non-negligible modification to Higgs couplings~\cite{Blum:2012kn, Randall:2007as}. However, these corrections are fully accounted for in Eqs.~(\ref{eq:rs2hdm}-\ref{eq:parr}), by assuming renormalized couplings in the potential~(\ref{eq:V2hdm}). 

For the purpose of understanding the phenomenology of specific models it is useful to express $r_b$ in terms of parameters in Eq.~(\ref{eq:V2hdm}). As a simple but interesting scenario, consider $\tan\beta\ge 3$, where we can use $(1/\tan\beta)$ as an expansion parameter~\cite{Blum:2012kn, Randall:2007as}. We assume some hierarchy between the masses of the two doublets\footnote{More precisely, we must require $B^2/(m_H^2-m_h^2)^2\ll1$, where $m_{H,h}$ are the mass eigenvalues. This condition can be put as $\tan\beta\gg\left(1-m_h^2/m_H^2\right)^{-1}$.}, $m_1^2>m_2^2$, and neglect CP-violation and loop corrections from charged and pseudo-scalar Higgs states. 
Defining the quantities
\beq\label{eq:defMB} M_1^2=m_1^2+\frac{\lambda_{35}h_2^2}{2}\,,\;\;\;B=m_{12}^2+\frac{\lambda_7h_2^2}{2},
\eeq
where $\lambda_{35}=\lambda_3+\lambda_5$\footnote{Compared with the basis of~\cite{Gunion:2002zf}, $\langle B/M_1^2\rangle \sim1/\tan\beta$ and our $\lambda_{35}$ equals their $\lambda_{345}$.},
a direct diagrammatic evaluation, treating the parameter $\langle B/M_1^2 \rangle=(1/\tan\beta)+\mathcal{O} (1/\tan^2\beta)$ as perturbation, yields
\beq\label{eq:rbb}
r_b&=&\left(1-\frac{m_h^2}{m_H^2}\right)^{-1}\left(1-\frac{1}{1+\Delta_b\tan\beta}\left(\frac{\lambda_{35}v^2}{m_H^2-m_h^2}-\frac{\lambda_7v^2}{m_H^2}\tan\beta+\frac{m_h^2}{m_H^2}\Delta_b\tan\beta\right)\right)\times\left\{1+\mathcal{O}\left(\frac{1}{\tan^2\beta}\right)\right\},\no \\
\eeq
where we used $m_H^2=\langle M_1^2 \rangle+\mathcal{O} (1/\tan^2\beta)$. The result for the coupling $r_\tau$ is similar to $r_b$, replacing $\Delta_b \to \Delta_\tau$. 

The approximation Eq.~(\ref{eq:rbb}) is useful because it allows us to understand generic predictions for a rather wide range of models~\cite{Blum:2012kn, Randall:2007as}. We will show in the next section that it is also fairly accurate. 
The couplings $\lambda_5, \lambda_7$ correspond to hard breaking of a $U(1)_{PQ}$ symmetry under which $(H_uH_d)$ is charged. Many phenomenologically relevant models (e.g. the MSSM) break $U(1)_{PQ}$ only softly and/or only at loop level, and thus $\lambda_{5,7}$ are suppressed while $\lambda_3$ can be $\mathcal{O}(1)$. Hence in many cases, $\lambda_{35}$ controls the correction $r_b$ and its sign determines whether $r_b$ is increased or decreased from unity. Some generic classes of SUSY models predict fixed signs of $\lambda_{35}$ and thus the direction of the shift in $r_b$. 

In the next subsections we apply Eqs.~(\ref{eq:parr}) and~(\ref{eq:rbb}) to SUSY models, including the MSSM and then extensions that can naturally accommodate $m_h=125$ GeV. We show that even though $\mathcal{O}(1)$ variations in $r_b$ can easily occur in such models, nevertheless significant predictive power is maintained, particularly in the approximate $PQ$ limit.

\subsection{MSSM analysis}
\label{sec:mssm}

The tree-level MSSM quartic Higgs potential is given by $G\equiv SU(2)_W\times U(1)_Y$ $D$-terms,
\[V_D=\sum_G\frac{g_G^2}{2}\left(H_u^\dag T_G^aH_u+H_d^\dag T_G^aH_d\right)^2=\frac{g^2+g'^2}{8}\left(\left|h_u^0\right|^2-\left|h_d^0\right|^2\right)^2+\cdots,\] where the dots represent the charged Higgs potential. $g$ and $g^\prime$ are the $SU(2)_W$ and the $U(1)_Y$ gauge couplings.
Mapping onto the 2HDM, we have at tree level
\beq \label{eq:tree}
\lambda_1=\lambda_2=-\lambda_{35}=\frac{g^2+g^{\prime 2}}{4}\approx0.14, \quad \lambda_5=\lambda_6=\lambda_7 = 0, \quad \Delta_b=\Delta_\tau=0. 
\eeq
The coupling $\lambda_{35}$ is negative, and so Eq.~(\ref{eq:rbb}) tells us that the value of $r_b$ is enhanced in the MSSM: $r_b\approx1+(m_Z^2+m_h^2)/m_H^2\approx1+0.25\left(m_H/300\,{\rm GeV}\right)^{-2}$. 

\begin{figure}[!h]
\begin{center}
\includegraphics[width=0.45\textwidth]{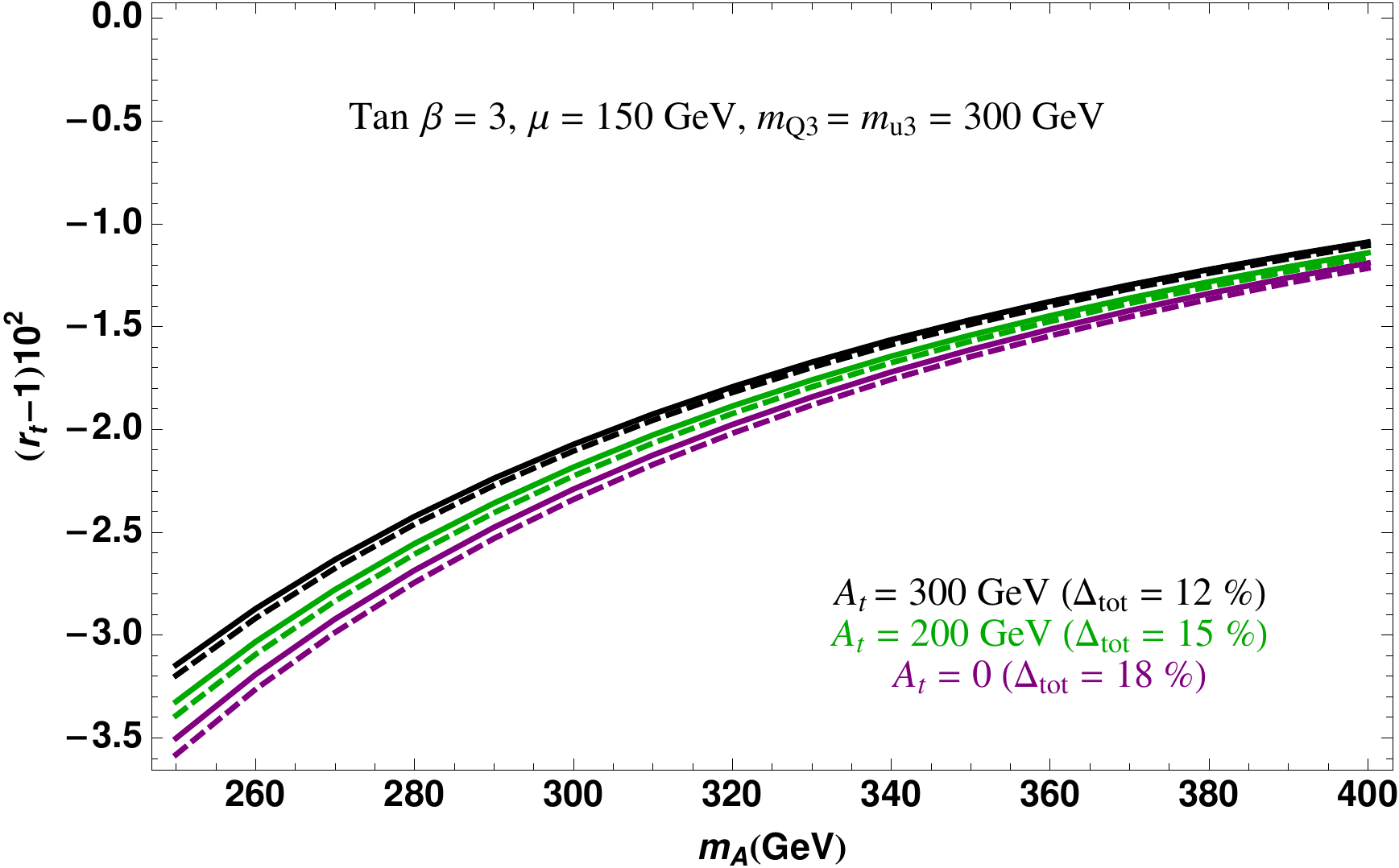}\quad
\includegraphics[width=0.4425\textwidth]{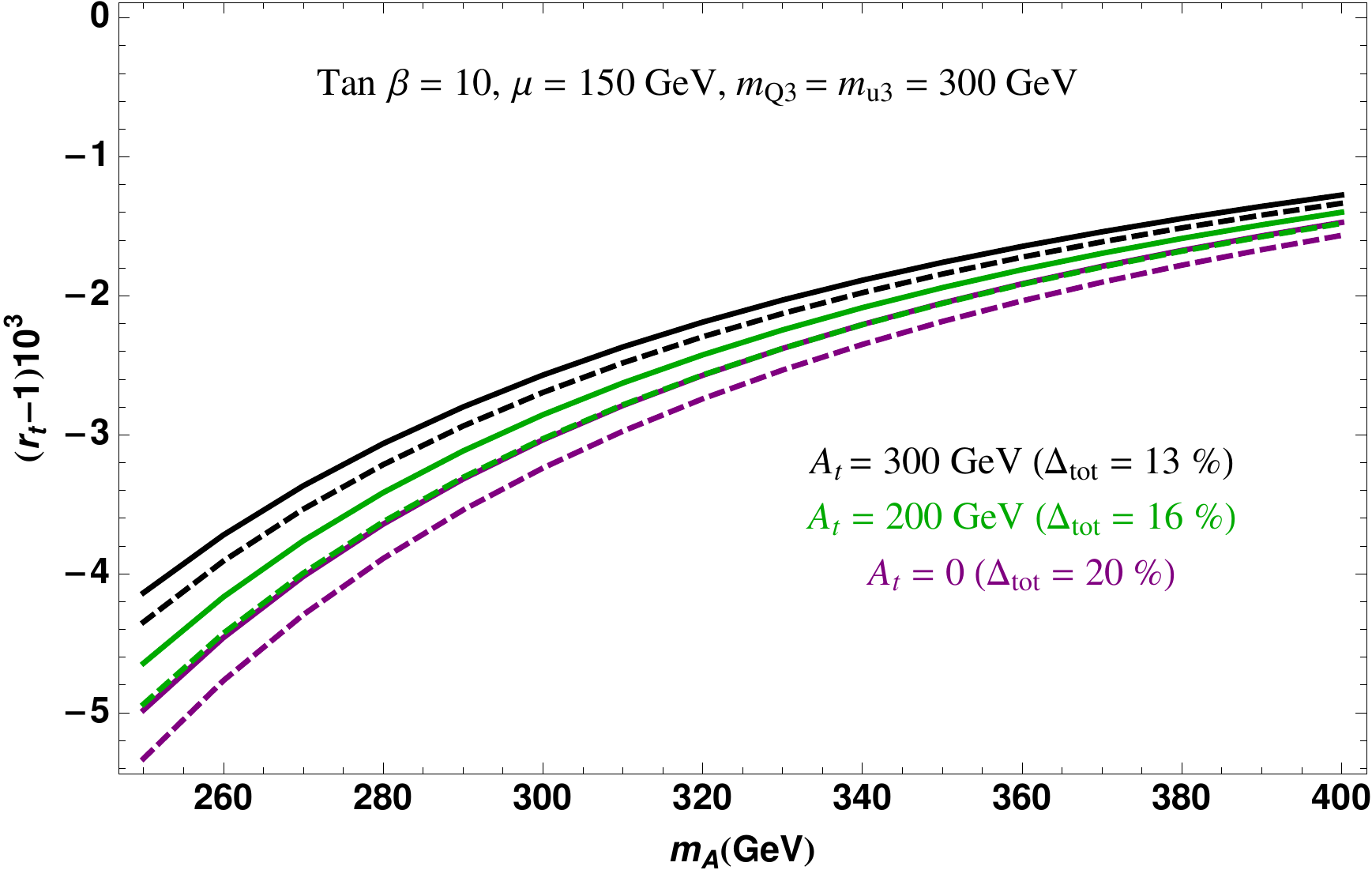} \\ \end{center}
\caption{$(r_t-1)$ as a function of $m_A$ for $\tan\beta = 3$ (left) and $\tan\beta = 10$ (right).  
In addition to the stop-higgsino loop with parameters $A_t$ and $\mu$ as shown, a sbottom-gluino contribution is also included with $m_{D3}=300$ GeV, $M_{\tilde g}=800$ GeV. The solid curves are derived by using the value of $r_b$, extracted from FeynHiggs, in our formula Eq.~(\ref{eq:parr}). The dashed curves are derived using FeynHiggs. Similar results are found for $r_V$.}
\label{fig:parr}
\end{figure}

Finite $\Delta_{b,\tau}$ and $\lambda_{5,6,7}$ arise at loop order; using the results of~\cite{Carena:1995bx, Hall:1993gn} we have, parametrically\footnote{A contribution to $\lambda_7$, coming from stau loops, may become comparable to the stop contribution if one allows very large $(A_\tau\mu/m_{\tilde\tau}^2)$.}
\beq\label{eq:numlam}
\lambda_7\sim-10^{-2}\left(\frac{A_t\mu}{m_{\tilde t}^2}\right),\;\;\;
\Delta_b\sim3\times10^{-3}\left[\left(\frac{A_t\mu}{m_{\tilde t}^2}\right)+4\left(\frac{m_{\tilde g}\mu}{\,m_{\tilde b}^2}\right)\right],\;\;\;
\Delta_\tau\sim10^{-4}\left(\frac{m_{\tilde B}\mu}{m_{\tilde \tau}^2}\right).
\eeq
We can now get back to estimate the validity of some of our previous approximations. First, while Eq.~(\ref{eq:parr}) automatically accounts for finite $\lambda_{5,6,7}$ (either radiatively generated or otherwise), it does not encode finite $\Delta_b$ and we should verify its aplicability. Note that $\Delta_b$ enters Higgs effective couplings multiplied by $\tan\beta$. However, from Eq.~(\ref{eq:numlam}) we expect that Eq.~(\ref{eq:parr}) should apply to reasonable accuracy even for $\tan\beta\sim10$, in which case, because of the relative $\tan^2\beta$ suppression, deviations in $r_t,\,r_V$ are already at most marginally relevant experimentally. In Fig.~\ref{fig:parr} we verify this point directly by comparing Eq.~(\ref{eq:parr}) with the full one-loop results from FeynHiggs. We find that for moderate $\tan\beta=3$ the agreement is excellent, while for $\tan\beta=10$ the agreement is better than 10\%. This makes  the overall error, incurred by using Eq.~(\ref{eq:parr}) for Higgs observables, no larger than order percent.  

\begin{figure}[!h]\begin{center}
\includegraphics[width=0.45\textwidth]{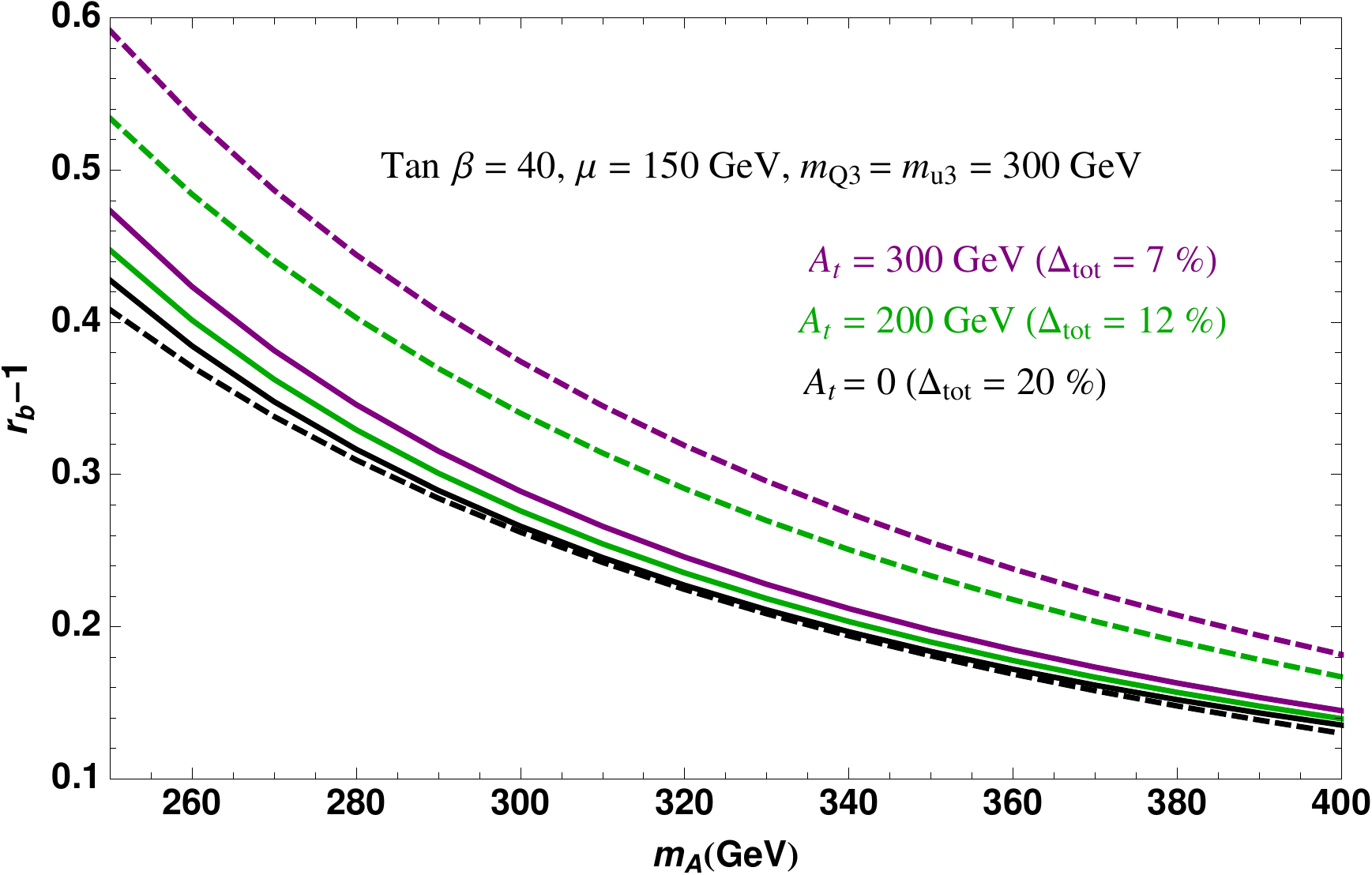}\quad
\includegraphics[width=0.45\textwidth]{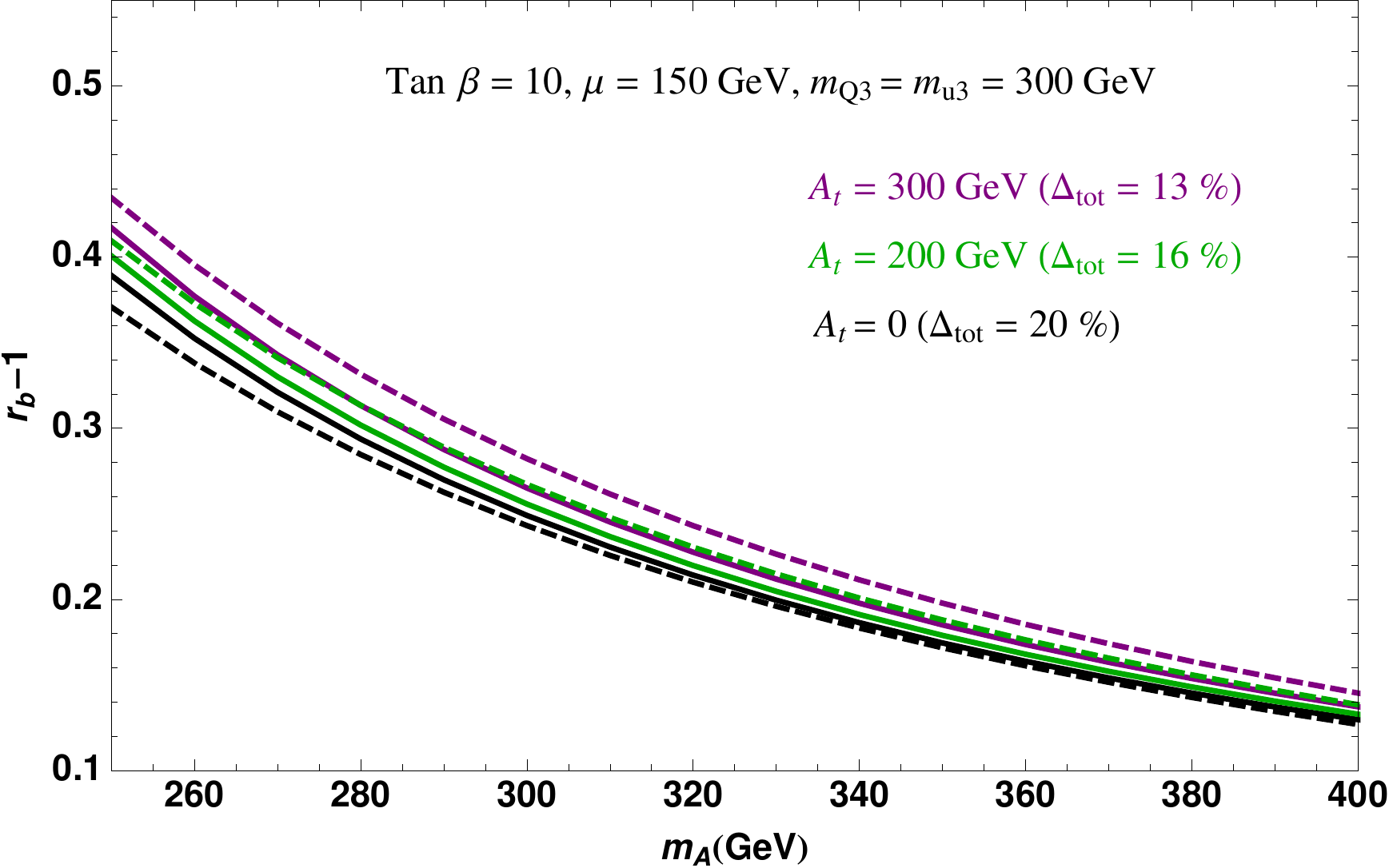} \\ \end{center}
\begin{center}
\includegraphics[width=0.45\textwidth]{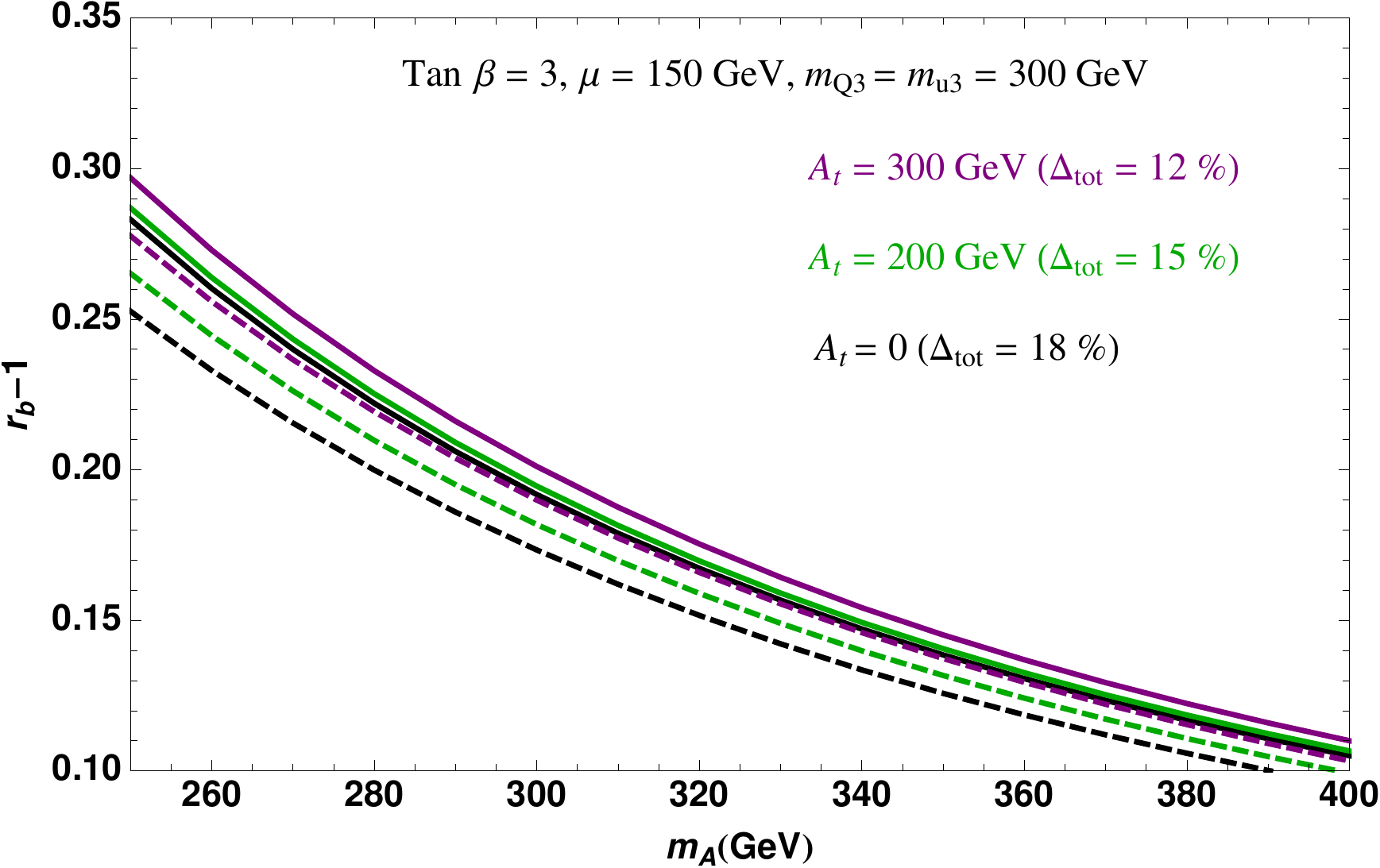}  \end{center}
\caption{$r_b-1$ as a function of $m_A$ for $\tan\beta = 40$ (upper left), $\tan\beta = 10$ (upper right), $\tan\beta = 3$ (lower). The solid curves are tree-level leading results in the large $\tan\beta$ approximation obtained from Eq.~(\ref{eq:rbb}) (setting $\Delta_b$ and $\lambda_7$ to be zero). The dashed curves are the full results computed with FeynHiggs. We decouple the right-handed sbottom in the computation.}
\label{fig:rb}
\end{figure}

Next, we consider the approximate $PQ$ result, Eq.~(\ref{eq:rbb}).  
As we argued in Sec. 2, naturalness favors $A_t \lesssim 300$ GeV.  
Thus unless $\tan\beta$ is very large, the contributions from $\lambda_7$ (and $\Delta_b$, $\Delta_\tau$) are subdominant. 
In Fig.~\ref{fig:rb} we study the accuracy of Eq.~(\ref{eq:rbb}) using the tree-level couplings~(\ref{eq:tree}) and comparing to the numerical results from FeynHiggs. We find that Eq.~(\ref{eq:rbb}) captures the correct result to $5\%$ accuracy for intermediate $\tan\beta=10$, and to better than $20\%$ accuracy for large (small) $\tan\beta=40\,(3)$, where $\tan\beta$-enhanced loop corrections (terms $\mathcal{O}(1/\tan^2\beta)$) are in effect. 

Last, we question the approximation $r_b\approx r_\tau$, that is violated by $\Delta_b\neq\Delta_\tau$. Again, the non-holomorphic corrections are small unless $\tan\beta$ is large, so we can use Eq.~(\ref{eq:rbb}). This gives
\beq\frac{r_b}{r_\tau}\approx\frac{r_{b0}+(1-r_{b0})\Delta_b\tan\beta}{r_{b0}+(1-r_{b0})\Delta_\tau\tan\beta},\eeq
with $r_{b0}=r_{\tau0}$ denoting the result for $\Delta_b=\Delta_\tau=0$. We conclude that $r_b\approx r_\tau$ is a reasonable approximation as long as $r_b,\,r_\tau=\mathcal{O}(1)$. The approximation can fail if the $hb\bar b$ and $h\tau\bar\tau$ couplings are strongly suppressed compared to the SM prediction; however, in that case we do not expect conclusive experimental information on these channels to become available any time soon.

\subsection{Non-decoupling $D$-term models}
\label{sec:Dterm}
In this section we study models with new gauge interactions, in which $D$-terms contribute the leading effect in raising the Higgs quartic couplings. Examples in the literature include~\cite{Batra:2003nj,Maloney:2004rc,Craig:2012hc}. We focus here on models in which the Higgs fields transform in a vector representation $(H_u,H_d)$ under the new gauge group, so that a $\mu$ term, $\mu H_u H_d$, is allowed in the superpotential; we show that these models generically predict an enhancement in the $hb\bar b$ and $h\tau\bar\tau$ couplings. Models that go beyond the vectorial charge assignment for the Higgs fields must gauge the $U(1)_{PQ}$ symmetry and tend to consist a hybrid of $D$-term and $F$-term models~\cite{Cvetic:1997ky,Morrissey:2005uz} that we discuss in the following section.

\begin{figure}[!h]\begin{center}
\includegraphics[width=0.45\textwidth]{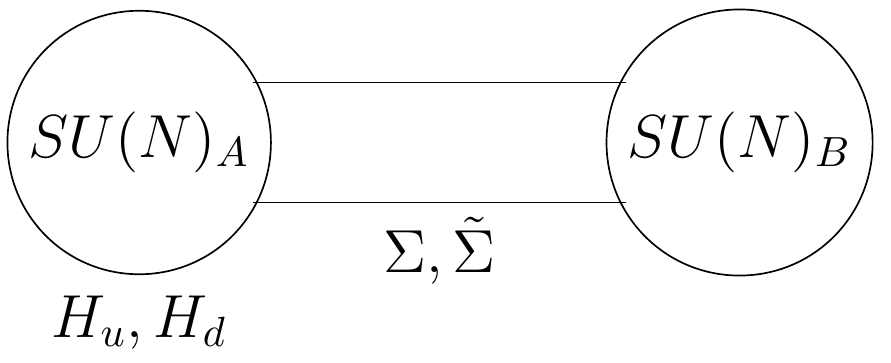}\  \end{center}
\caption{The fundamental building block in vector-like $D$-term models.}
\label{fig:moose}
\end{figure}

Consider two-site Moose models with a product gauge group $SU(N)_A \times SU(N)_B$ and gauge couplings $g_A$ and $g_B$ (for example, $SU(N) = SU(2)$ in~\cite{Batra:2003nj}), depicted in Fig.~\ref{fig:moose}. The product group is broken to the diagonal, identified as the SM electroweak gauge group $G$. 
This can be done, for instance, by introducing bi-fundamental link fields ($\Sigma$, $\tilde{\Sigma}$), developing VEVs $\langle\Sigma\rangle=\langle\tilde{\Sigma}\rangle$ through a superpotential $W=\lambda({\rm det} \Sigma + {\rm det} \tilde{\Sigma})+f \,{\rm Tr} \Sigma\tilde{\Sigma}$. The broken generator vector multiplets acquire mass $M_V$. To arrange for a non-decoupling effect, large soft SUSY breaking mass $M_s$ can be introduced for the link fields. Then, the low energy effective potential of these models takes the same form as the MSSM $D$-term potential with rescaled coefficients:
\beq \label{eq:D-potential}
V_D=\sum_G\frac{g_G^2}{2}\left(1+\frac{g_A^2}{g_B^2}\frac{M_s^2}{M_V^2+M_s^2}\right)\left(H_u^\dag T_G^aH_u+H_d^\dag T_G^aH_d\right)^2\supset\frac{g^2(1+\Delta)+g'^2(1+\Delta^\prime)}{8}\left(\left|h_u^0\right|^2-\left|h_d^0\right|^2\right)^2.\no \\
\eeq
Further details of the derivation leading to Eq.~(\ref{eq:D-potential}) can be found in App.~\ref{app:D}. 

The effect on the Higgs quartic potential is simply:
\beq
-\lambda_1=-\lambda_2=\lambda_{35}=\lambda_{35}^{\rm MSSM}\left(1+\frac{g^2\Delta+g'^2\Delta'}{g^2+g'^2}\right).
\eeq
The key point is that the sign of $\lambda_{35}$ is always {\em negative}, as long as $H_u, H_d$ are assigned charges in a vector-like representation. Therefore, by Eq.~(\ref{eq:rbb}), these models generically predict an enhancement of $r_b$. In particular, for light stops and $\tan\beta>$ few, we can approximate $\lambda_{35} v^2 \approx -\lambda_2^2v^2 \approx -m_h^2$. In the absence of hard PQ-breaking, $\lambda_7\ll(\lambda_{35}/\tan\beta)$, Eq.~(\ref{eq:rbb}) is then simplified to 
\beq
r_b \approx \left(1-\frac{m_h^2}{m_H^2}\right)^{-2}.
\label{eq:rbd}
\eeq

\begin{figure}[!h]\begin{center}
\includegraphics[width=0.45\textwidth]{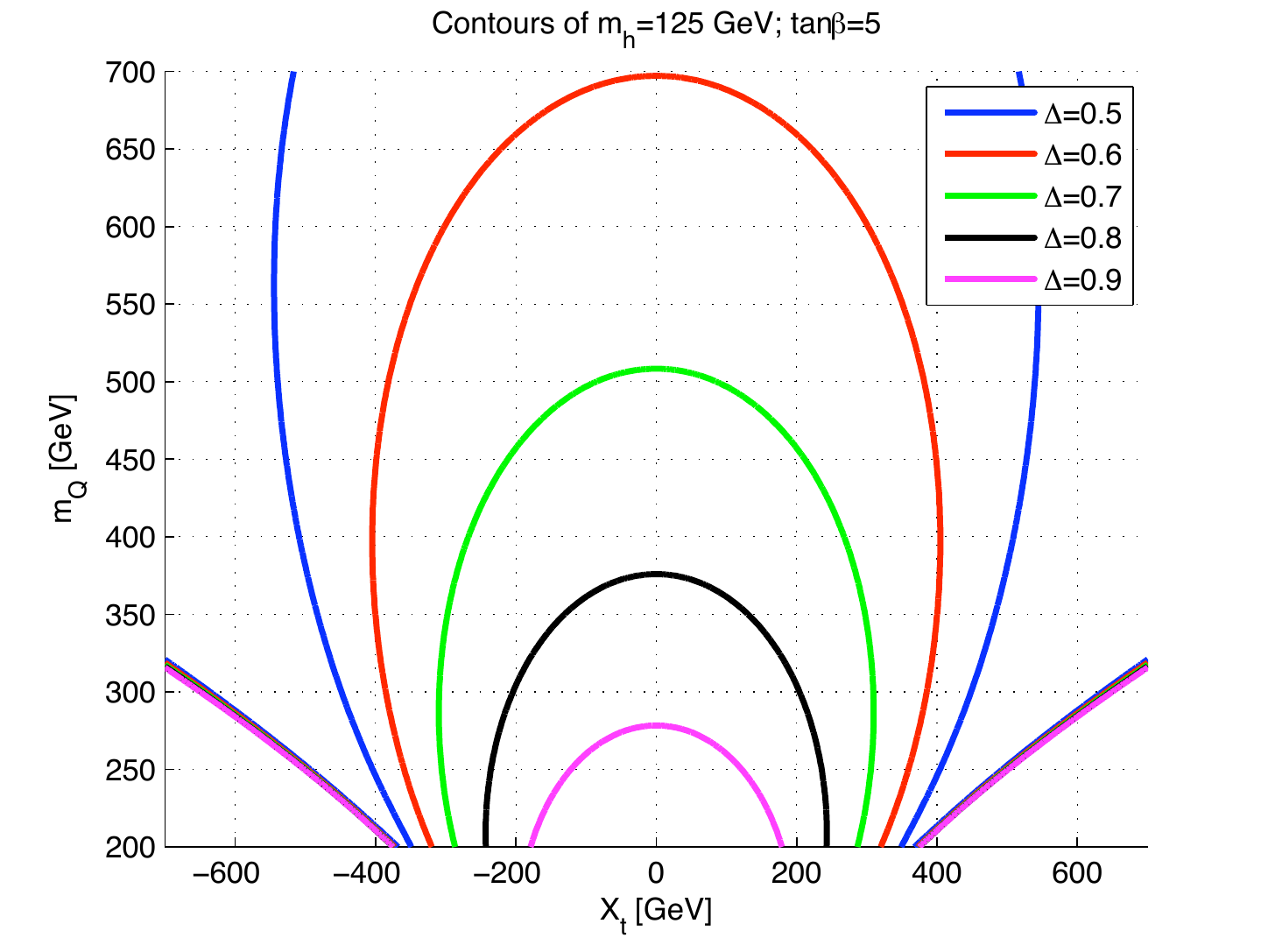}\quad
\includegraphics[width=0.45\textwidth]{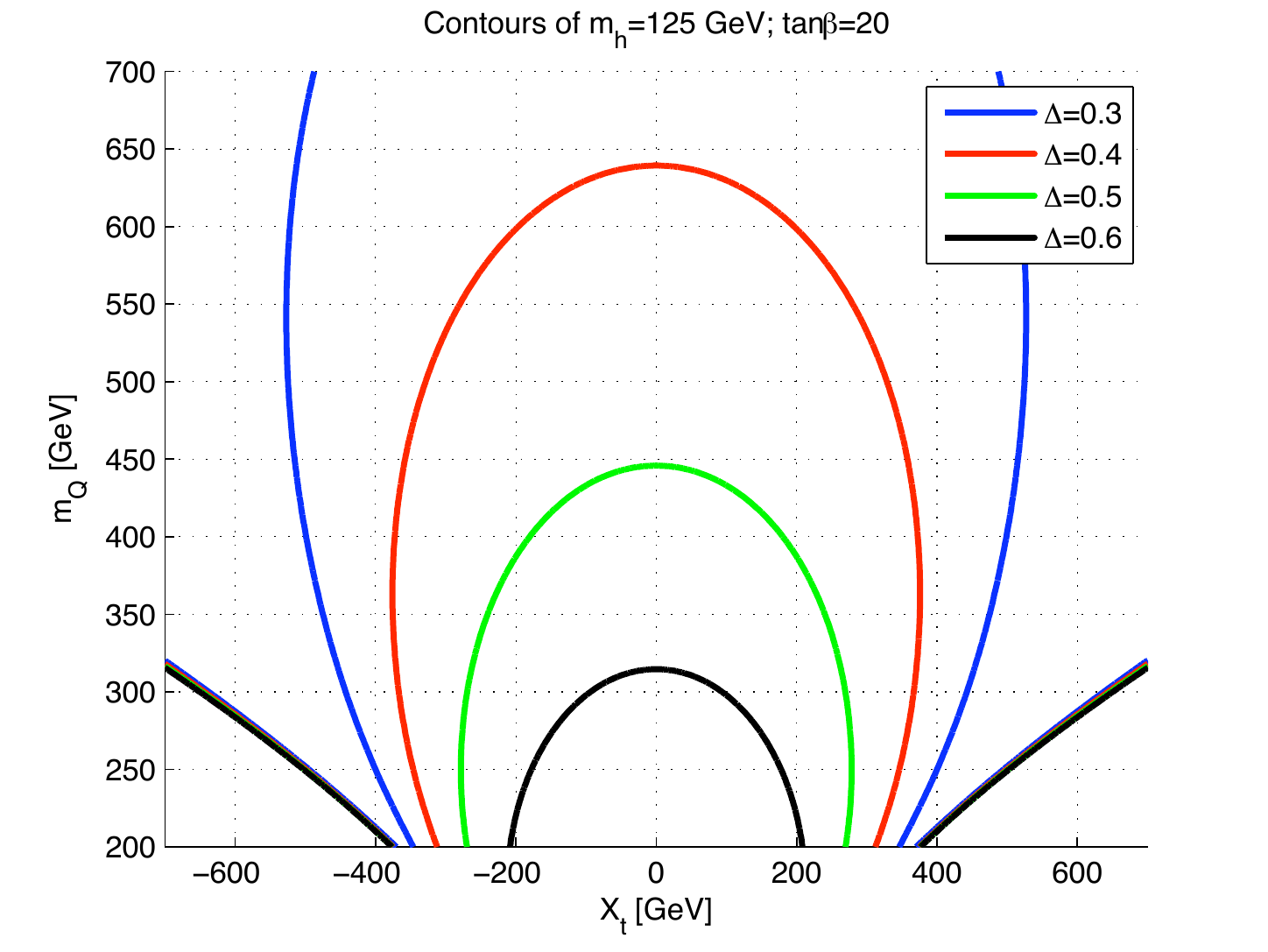}  \end{center}
\caption{Contours of $m_h=125$ GeV in the $(X_t,m_{Q_3})$ plane, for different values of the $D$-term correction parameter $\Delta$. Left: $\tan\beta=5$; Right: $\tan\beta=20$. We have set $\Delta=\Delta'$ for simplicity, but the results are not sensitive to this choice.}
\label{fig:D}
\end{figure}

More specifically, accepting $m_h=125$ GeV, we can compute the size of the required $\Delta$. In Fig.~\ref{fig:D} we set $\Delta=\Delta'$ and plot contours of $m_h=125$ GeV for different values of $\Delta$ in the stop soft mass-mixing plane, using Eq.~(\ref{eq:mhst}) modified by the $D$-term correction. Requiring light stops with fine-tuning no worse than 1:10, we find:  
\beq\label{eq:D05}\Delta\gsim0.5,\eeq 
giving a deviation, $(r_b-1)$, smaller by $10\%$ than the value obtained from Eq.~(\ref{eq:rbd}). 
As we comment in App.~\ref{app:D}, Eq.~(\ref{eq:D05}) implies that $D$-term models that address $m_h=125$ GeV should indeed be close to the non-decoupling regime.  

Note that the $D$-term corrections lead to a slight numerical modification in our estimate of the $Z$ mass fine-tuning, where, roughly, we should replace $m_Z\to(1+\Delta) m_Z$ in Eq.~(\ref{eq:Dz}). Compared to the usual MSSM estimate, this allows for somewhat heavier stops, $m_{\tilde t}\sim500$ GeV, to still be consistent with tuning of order 1:10. This has a small effect on our results concerning the viable range for stop loop effects in $h\gamma\gamma$ and $hGG$, allowing a slightly more pronounced {\it decrease} in $r_G^{\tilde t}$ (increase in $r_\gamma^{\tilde t}$).

\subsection{Non-decoupling $F$-term models}\label{sec:Fterm}

\begin{figure}[!h]\begin{center}
\includegraphics[width=0.6\textwidth]{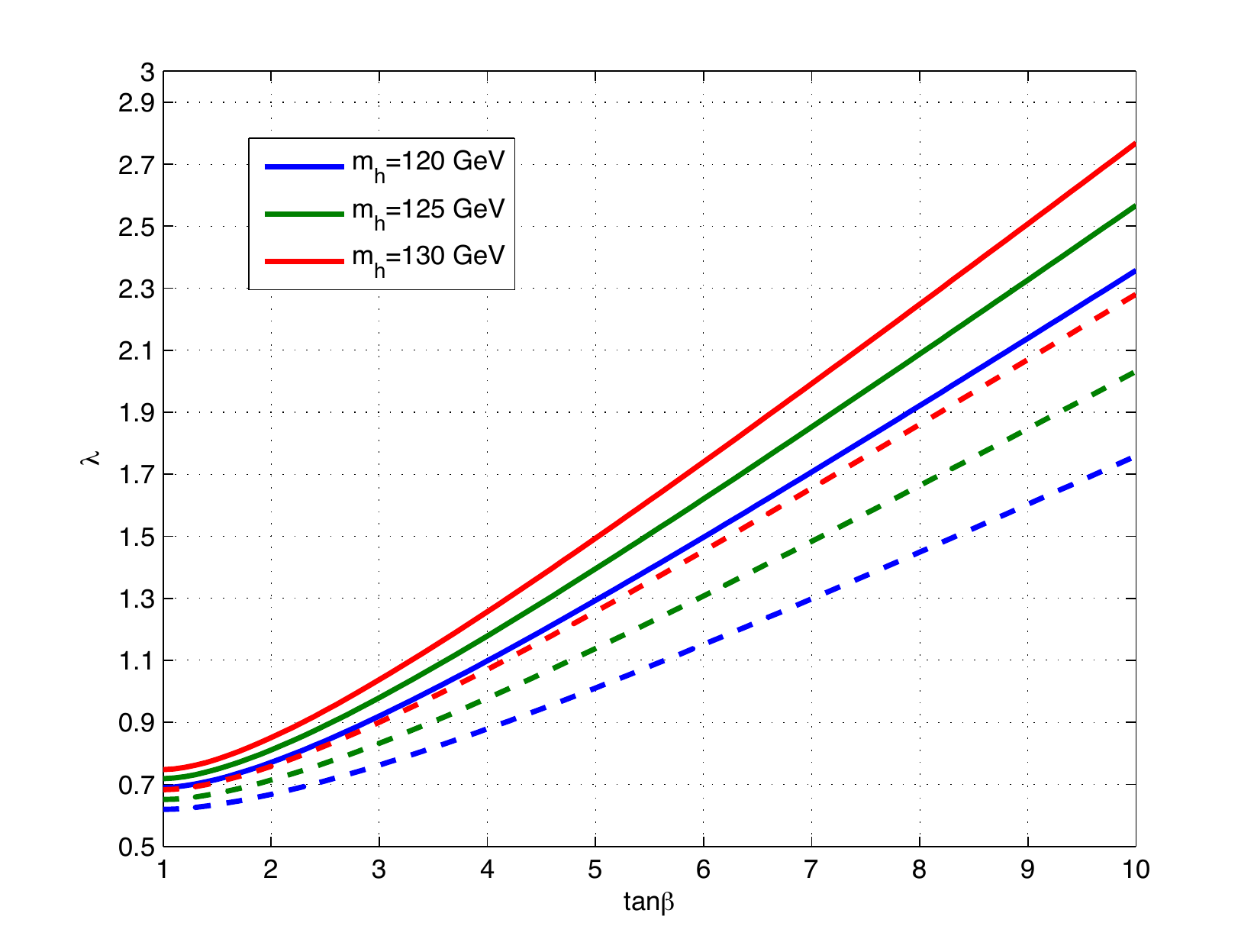}\end{center}
\caption{Contours of $m_h=120,125,130$ GeV, in the $(\tan\beta,\lambda)$ plane. Smooth lines: tree level; dashed lines: including stop correction with $X_t=0,\,m_{\tilde t_1}=m_{\tilde t_2}=380$ GeV.}
\label{fig:lambdaSUSY}
\end{figure}

Here we consider models that raise the Higgs mass through new interactions in the superpotential. The classic example includes a SM singlet, interacting with the Higgs doublets via 
\beq\label{eq:S} \delta\WW=\lambda SH_uH_d.\eeq
(We will get the same results by making $S$ a hypercharge-neutral SU(2) triplet.)
If $S$ is given a large soft SUSY breaking mass, $m_s^2\gg m_H^2,M^2_s$, with $m_H$ being the mass of the heavier MSSM Higgs doublet and $M_s$ a possible supersymmetric mass for $S$, then the effective potential below $m_s$ is modified with a non-decoupling correction,
\beq V=V^{\rm MSSM}+|\lambda|^2|H_uH_d|^2.\eeq
This gives, in our notation of Eq.~(\ref{eq:V2hdm}),
\beq
\lambda_4=\lambda_4^{\rm MSSM}-|\lambda|^2, \quad \lambda_{35}=\lambda_{35}^{\rm MSSM}+|\lambda|^2.
\eeq
By Eq.~(\ref{eq:rbb}), these models tend to decrease $r_b$. To estimate the size of the effect, note that the correction to the Higgs mass, still neglecting mixing, is
\beq\label{eq:dmhlamsusy}\delta m_h^2=m_Z^2\left(\frac{2|\lambda|^2}{g^2+g'^2}\right)\sin^22\beta.\eeq
In Fig.~\ref{fig:lambdaSUSY} we plot the value of $\lambda$ required for $m_h=125$ GeV, by adding Eq.~(\ref{eq:dmhlamsusy}) to Eq.~(\ref{eq:mhst}). 
Mixing with the heavy singlet $S$ reduces the lightest Higgs mass due to level splitting~\cite{Hall:2011aa}; as a result, for fixed $m_h$, the value of $\lambda$ in Fig.~\ref{fig:lambdaSUSY} serves only as a lower bound. We conclude that $\lambda_{35}$ is always positive in this model and much larger than its gauge-coupling value in the MSSM. This reduces the $hb\bar b$ coupling below its SM value.

As pointed out in~\cite{Hall:2011aa}, the non-decoupling limit discussed above has limited applicability because naturalness constrains $m_s\lsim$TeV. In contrast with $D$-term models, however, where electroweak precision tests constrain the SUSY scale, the singlet $F$-term example is phenomenologicallly viable also in the SUSY limit. In analogy with the $D$-term example, adding a supersymmetric mass term, $\delta\WW\supset(M_s/2)S^2$, the shift in $\lambda_{35}$ is suppressed by factors of $(m_s/M_s)$; however, a supersymmetric correction $\lambda_7=-(\lambda^2\mu^*/M_s)$ is generated. The $\lambda_7$ term modifies the Higgs mass by $\delta m_h^2\propto-(\lambda_7 v^2/\tan\beta)$. With $|M_s| = 700$ GeV, $|\mu| = 200$ GeV, $\tan\beta = 2$ and $|\lambda| = 0.7$, the Higgs mass could be raised to 125 GeV~\cite{Dine:2007xi,Blum:2009na} with just a little help from $\sim$300 GeV stops. From Eq.~(\ref{eq:rbb}), the small negative $\lambda_7$ will also act to decrease $r_b$. 

We should stress, however, that it is not difficult to construct $F$-term models that do not decrease $r_b$. In particular, models that attempt to produce the $\mu$ term dynamically via a weak-scale singlet vacuum expectation value, tend to predict a light singlet, in which case the 2HDM analysis ceases to apply. An example is the Z3-NMSSM,
\beq\label{eq:Fm2}\WW=\lambda SH_uH_d+\frac{\kappa}{3}S^3.\eeq
Parameteric scans of this model show that the shift in $r_b$ does not have a definite direction once $S$ is allowed to be light~\cite{Ellwanger:2012ke, King:2012is}. To understand how this can happen, consider Eq.~(\ref{eq:Fm2}) with the following hierarchy of masses: $m_H^2\approx m_{H_d}^2 > |m_s|^2 > |m_{H_u}|^2$, where $m_s^2<0$. Integrating out first $H_d$ and then $S$, and expanding to leading order in $(1/\tan\beta)$, 
we have
\beq \label{eq:rbf}
r_b \approx 1-\frac{|\lambda|^2 v^2}{|m_s^2|}+\frac{ v^2}{m_H^2-m_h^2}\left(\frac{g^2+g^{\prime2}}{4}+\frac{|\lambda|^4}{2|\kappa|^2}\right).
\eeq
Notice that there is no longer a correction scaling as $(|\lambda|^2 v^2/m_H^2)$; instead there are two opposite-sign contributions that could be parametrically comparable. 

Another example that increases the Higgs mass with no definite effect in the $hb\bar b$ coupling is found by adding an SU(2) triplet chiral superfield, $\Delta_{-}\sim(1,3)_{-1}$, with super potential $\delta\WW=\lambda\Delta_{-}H_uH_u$. With a large SUSY breaking mass $m_{s-}$, integrating out the field $\Delta_{-}$ inserts a hard SUSY-breaking correction $\lambda_2\to\lambda_2+|\lambda|^2$ without modifying any of the other quartics in Eq.~(\ref{eq:V2hdm}), lifting the Higgs mass with no further effect on the Higgs-fermion couplings.

Our main conclusions from the discussion in Secs.~\ref{sec:Dterm} and~\ref{sec:Fterm} are these: (i) $\mathcal{O}(1)$ modifications to $r_{b,\tau}$ are plausible through Higgs mixing in concrete extensions of the MSSM, that address the Higgs mass; (ii) measuring a deviation in $r_b$ will have strong implications for generic classes of models, in particular, non-decoupling $D$-term and $F$-term models, that predict opposite sign effects. The value of $(r_b-1)$ will provide very suggestive hints for the masses, quantum numbers and couplings of new particles beyond the MSSM.

\section{Natural SUSY predicts}\label{sec:data}

A $125$ GeV SM-like Higgs has several accessible decay and production modes. However, at this stage experimental uncertainties at the LHC and Tevatron are large. It is reasonable to estimate that even with the LHC 14 TeV run well under way ($L_{\mathrm{int}}=30\;\mathrm{fb}^{-1}$), individual couplings will only be measured to $20\% - 40\%$ accuracy~\cite{Klute:2012pu}. These estimates are naive as far as systematics are concerned, but they should fall in the right ballpark. Given the experimental prospects, we limit the discussion to theoretical predictions of Higgs couplings that are valid to about $5\%$. This translates to $\sim10\%$ accuracy in production and decay rates. Natural SUSY is perfectly capable of inducing much larger deviations, in which case our analysis will help to discriminate between different models.

Following the discussion in Secs.~\ref{sec:loop} and~\ref{sec:hmix}, we classify the contributions to modified Higgs couplings into loop effects and mixing. We can now parametrize Higgs observables using four independent variables, two for each class of effects. These variables are
\beq\label{eq:4var}\tan\beta,\;\;\;r_b,\;\;\;\left(0.85<r_G^{\tilde t}<1.5\right),\;\;\;\left(0.7<r_\gamma^{\tilde\chi^\pm}<1.1\right).\eeq
Using $r_b$ and $\tan\beta$ in Eq.~(\ref{eq:parr}) we can compute $r_t$ and $r_V$. Using $r_G^{\tilde t},\,r_\gamma^{\tilde\chi^\pm},\,r_t$ and $r_V$ we can compute the observable factors $r_G$ and $r_\gamma$. A set of four free parameters, with a limited range of values, is a rather predictive framework considering that experimental Higgs analyses will be sensitive to $\mathcal{O}(10)$ different production/decay channels.
A few comments are in order:
\begin{itemize}
\item The fact that four variables suffice to describe Higgs production and decay is not special to our SUSY framework, but simply the result of assuming a 2HDM at the weak scale. This assumption gave us $r_t$ and $r_V$ in terms of $r_b$ and $\tan\beta$,\footnote{Alternatively, of course, we could have used the angle variables $\beta,\,\alpha$.} while to describe the Higgs photon and gluon couplings we could have chosen to use $r_\gamma$ and $r_G$ directly.
\item Our choice of the $r_\gamma^{\tilde\chi^\pm}$ and $r_G^{\tilde t}$ variables, is based on our ability to predict the viable numerical ranges for them in the particular framework of natural SUSY. It is of interest to spell out what part of the constraints on $r_\gamma^{\tilde\chi^\pm}$ and $r_G^{\tilde t}$ actually comes from naturalness vs. experimental limits: 
\begin{itemize}
\item The upper limit $r_G^{\tilde t}<1.5$ comes from imposing the direct constraint $m_{\tilde t}>$100 GeV and, when it is saturated, stops contribute significantly to $(\Delta\rho/\rho)$. Requiring further that the stop-sbottom contribution to $(\Delta\rho/\rho)$ does not exceed $4\sigma$ would lower this bound to $r_G^{\tilde t}<1.3$. Similarly, both the upper and lower limits on the chargino contribution $r_\gamma^{\tilde\chi^\pm}$ do not involve naturalness considerations, but merely the direct constraint $m_{\tilde\chi^\pm}>94$ GeV.
\item The lower limit $r_G^{\tilde t}>0.85$ does arise from naturalness considerations. More specifically, it comes about by limiting the stop mixing to be modest. In some models, e.g. a singlet extension $F$-term model with sizable $\lambda$, the naturalness constraint may be relaxed somewhat and with it the lower bound on $r_G^{\tilde t}$. 
\item Finally, it was essentially naturalness (though assisted by direct constraints) that guided us to neglect the stau and sbottom loop corrections to $r_\gamma$ and $r_G$.
\end{itemize}
\end{itemize}

Let us apply our analysis to a number of experimental channels, defining the signal strength $\mu_X=\sigma\times BR(X)/{\rm SM}$. First, a 125 GeV SM Higgs has a partial width of $\approx64.4\%$ to $b\bar b$ and $\tau\bar\tau$, $24.3\%$ to $WW$ and $ZZ$, $8.5\%$ to gluon pairs and $2.7\%$ to $c\bar c$. For our purpose it suffices to approximate the total width modification by 
\beq\Gamma/\Gamma_{\rm SM}\equiv\mu_{\rm tot}\approx0.64r_b^2+0.24r_V^2+0.09r_G^2+0.03r_t^2.\eeq
Note that $\mu_{\rm tot}$ depends mostly on Higgs mixing through $r_b$ and $r_V(r_b, \tan\beta)$. 
Consider now the following six processes, with GF, VBF and AP standing for gluon fusion, vector boson fusion and associated production, respectively:
\beq&\mu_{\gamma\gamma;GF}=\frac{r_G^2r_\gamma^2}{\mu_{\rm tot}},\label{eq:gamgf}\\ 
&\mu_{\gamma\gamma;VBF}=\frac{r_V^2r_\gamma^2}{\mu_{\rm tot}},\label{eq:gamvbf}\\
&\mu_{ZZ,WW;GF}=\frac{r_G^2r_V^2}{\mu_{\rm tot}},\label{eq:vgf}\\
&\mu_{bb,\tau\tau;AP}=\frac{r_V^2r_b^2}{\mu_{\rm tot}}.\label{eq:bap}
\eeq
Some relevant questions, motivated in part by the current experimental situation (summarized in Sec.~\ref{ssec:curdat}), are the following:
\begin{enumerate}
\item {\bf What is the maximal enhancement for $h\to\gamma\gamma$ in GF production?}
The bound is obtained with large positive stop corrections, suppressed $hb\bar b$ coupling $r_b\ll1$ and SM-like $r_t\approx r_V\approx1$. We find
\beq\mu_{\gamma\gamma;GF}\lsim4.7\;{\rm to}\;6.2,\eeq 
where we allow stops to provide $r_G^{\tilde t}=$1.3 to 1.5, respectively. $\mu_{\gamma\gamma;GF}\sim2$ is easy to achieve in beyond-MSSM SUSY by suppressed $hb\bar b$ coupling, together with a little help from natural stops. Note that our upper limits on the chargino loop contribution implies that if the diphoton rate is increased by more than $\sim20\%$, then so should the $ZZ,WW$ rates. (See bullet (3) below).
\item {\bf What is the maximal enhancement for $h\to b\bar{b}$ in AP?} This question is partially motivated by the hint for factor $\sim2$ enhancement in $h\to b\bar{b}$ associated production, reported recently by the Tevatron experiments. A constraint comes simply from 2HDM trigonometry,
\beq\mu_{bb;AP}=\frac{r_V^2r_b^2}{\mu_{\rm tot}}<1.5.\eeq
We obtain the upper bound of $1.5$ by examining $0.1<\tan\beta<50$, $0<r_b<10$, and varying $r_G^{\tilde t}$ in the range of Eq.~(\ref{eq:4var}). The maximal value obtains for $\tan\beta\gsim30$ and $r_b\gsim3.5$. Experimentally establishing a nontrivial lower bound on $\mu_{bb;AP}$ would have profound implications for natural SUSY. For instance, establishing $\mu_{bb;AP}\geq1.4$ would provide the lower bounds $\tan\beta>8,\,r_b>2$, ruling out $F$-term models like $\lambda$SUSY, while making a strong case for $D$-term models that enhance $hb\bar b$. Interpreted within the latter models, by Eq.~(\ref{eq:rbd}) the measurement would imply a challenging (but still consistent with direct searches) upper bound on the heavier Higgs doublet, $m_H\lsim230$ GeV. This discussion applies also to $\mu_{\tau\tau;AP}$.
\item {\bf What is the maximal ratio of $h\to\gamma\gamma$ vs. $ZZ,WW$ in the GF production channel?} Both LHC experiments report $\mu_{\gamma\gamma;GF}\gsim3\times \mu_{WW;GF}$, and a low $WW$ rate is also found at the Tevatron. The ratio and the answer are given by
\beq\label{eq:gam2Z}\frac{\mu_{\gamma\gamma;GF}}{\mu_{WW,ZZ;GF}}=\frac{r_\gamma^2}{r_V^2}=\left|1.28-0.28(r_G^{\tilde t}r_t/r_V)+(\delta r_\gamma^{\tilde\chi^\pm}/r_V)\right|^2<1.4,\eeq
where $\delta r_\gamma^{\tilde\chi^\pm} \equiv r_\gamma^{\tilde\chi^\pm} -1$. 
The numerical value of 1.4 answers a slightly modified question, namely: ``{\bf what is the maximal value of Eq.~(\ref{eq:gam2Z}), assuming that $h\to\gamma\gamma$ is not suppressed}, i.e. assuming $\mu_{\gamma\gamma;GF}\geq0.75$?" This is more relevant, because to truly maximize Eq.~(\ref{eq:gam2Z}) one would need to take $r_V$ small, which would diminish the actual observed $\gamma\gamma$ rate. Imposing $\mu_{\gamma\gamma;GF}\geq0.75$ and varying the  variables in Eq.~(\ref{eq:4var}), we find the numerical bound in Eq.~(\ref{eq:gam2Z}).
\item {\bf What is the maximal ratio of $h\to\gamma\gamma$ in the VBF vs. the GF channels?} Again, we further impose $\mu_{\gamma\gamma;GF}\geq0.75$, this time in order to avoid irrelevant solutions with $r_t\ll1$ and vanishing GF production. In CMS, the reported VBF to GF ratio is $\sim2$. The upper bound we obtain is:
\beq\frac{\mu_{\gamma\gamma;VBF}}{\mu_{\gamma\gamma;GF}}=\frac{r_V^2}{r_G^2}<1.5.\eeq
\end{enumerate}

The sample questions above give a sense to the level of predictive power in natural SUSY. However, the usefulness of reducing the number of free parameters to just a few comes mostly in interpreting a larger set of independent measurements. In what follows we demonstrate this point by fitting the parameters in Eq.~(\ref{eq:4var}) to current data and interpreting the results.

\subsection{Interpreting current data}\label{ssec:curdat}

Here we inspect and interpret current results from the LHC and the Tevatron experiments. We consider the following channels:
\begin{enumerate}
\item CMS \cite{CMSphoton} and ATLAS \cite{ATLASphoton} $\gamma\gamma$, including the CMS dijet tagged category.
\item CMS \cite{CMSDiboson} and ATLAS \cite{ATLASDiboson} $ZZ$.
\item CMS \cite{Chatrchyan:2012ty} and ATLAS \cite{ATLASWW} $WW$.
\item CMS \cite{Chatrchyan:2012vp} and ATLAS \cite{atlastautau} $\tau \tau$.
\item CMS \cite{Chatrchyan:2012ww} and ATLAS \cite{atlasbb} $b \bar b$.
\item CDF \cite{CDFhiggscomb} and D0 \cite{D0higgscomb} combined $WW$ and $b \bar b$.
\end{enumerate}	
We included only the associated production mode for the $\tau \tau$ and $b \bar b$ channels and the $\gamma \gamma$ dijet tagged category of CMS was interpreted as containing a $30\%$ of events coming from GF and the remaining $70\%$ from VBF production. This is justified by the CMS estimate of the GF contamination, that can be found in \cite{CMSphoton}. The rates we use in this study were reported by the experiments as best fits to the signal strength and can be found in the Higgs combination papers and conference notes \cite{cmshiggscomb, atlashiggscomb, TEVNPH:2012ab}.

Our approach to interpreting the current data is as follows. First, we adopt a reference channel and assume some particular value for it, consistent with the experimental result. Then, assuming this reference value, we vary the parameters in Eq.~(\ref{eq:4var}) and obtain the viable ranges for all of the other experimental channels. We assume $m_h=125$ GeV, but our results have little sensitivity to varying $m_h$ by $\pm2$ GeV. 

\begin{figure}[!h]\begin{center}
\includegraphics[width=0.8\textwidth]{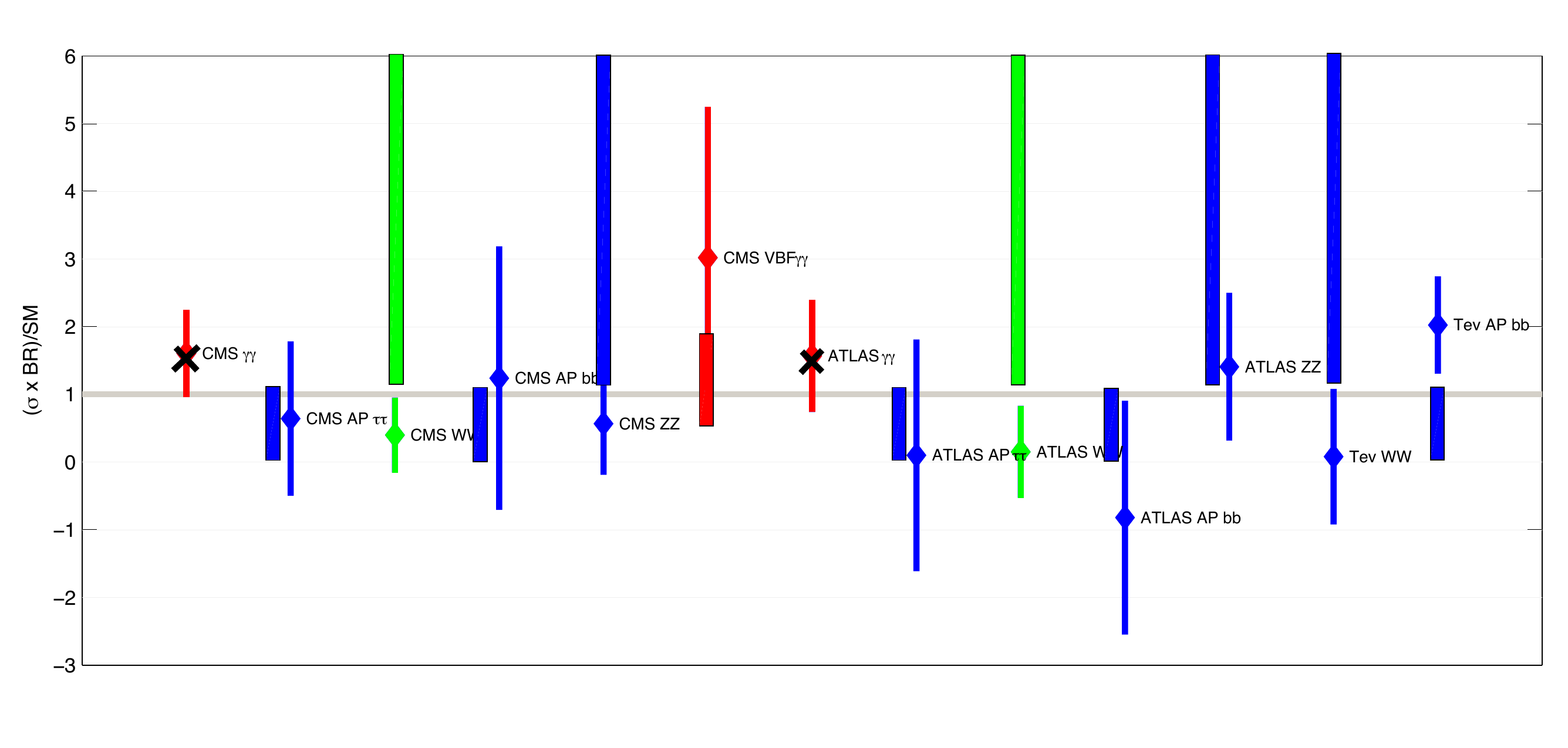} 
 \end{center}
\caption{Natural SUSY predictions, assuming GF production $h\to\gamma\gamma$ as indicated by the black crosses, are marked by thick bands.}
\label{fig:ratesGam}
\end{figure}

As the first reference channel, we select the LHC measurements of GF production $h\to\gamma\gamma$. CMS and ATLAS report consistent results for this channel, with a relatively small experimental error, $\left(\mu_{\gamma\gamma;GF}\right)^{\rm exp}\approx1.5\pm0.5$. As reference value we choose: $\left(\mu_{\gamma\gamma;GF}\right)^{\rm ref}=1.5$. The results are shown in Fig.~\ref{fig:ratesGam}. Within natural SUSY, our reference value for $\mu_{\gamma\gamma;GF}$ is in about $1\sigma$ tension with each of the reduced $WW$ and $ZZ$ at CMS, the reduced $WW$ at ATLAS, and the reduced (enhanced) $WW$ $(b\bar b)$ at the Tevatron. Natural SUSY predicts that the ratio between the $h\to\gamma\gamma$ and $h\to WW,ZZ$ rates should get closer to unity with future data. If this does not happen, then we will have an indication that some of our basic assumptions were wrong. A simple possibility would be e.g. that additional charged particles, beyond the MSSM matter content, contribute to the $h\gamma\gamma$ vertex.

\begin{figure}[!h]\begin{center}
\includegraphics[width=0.8\textwidth]{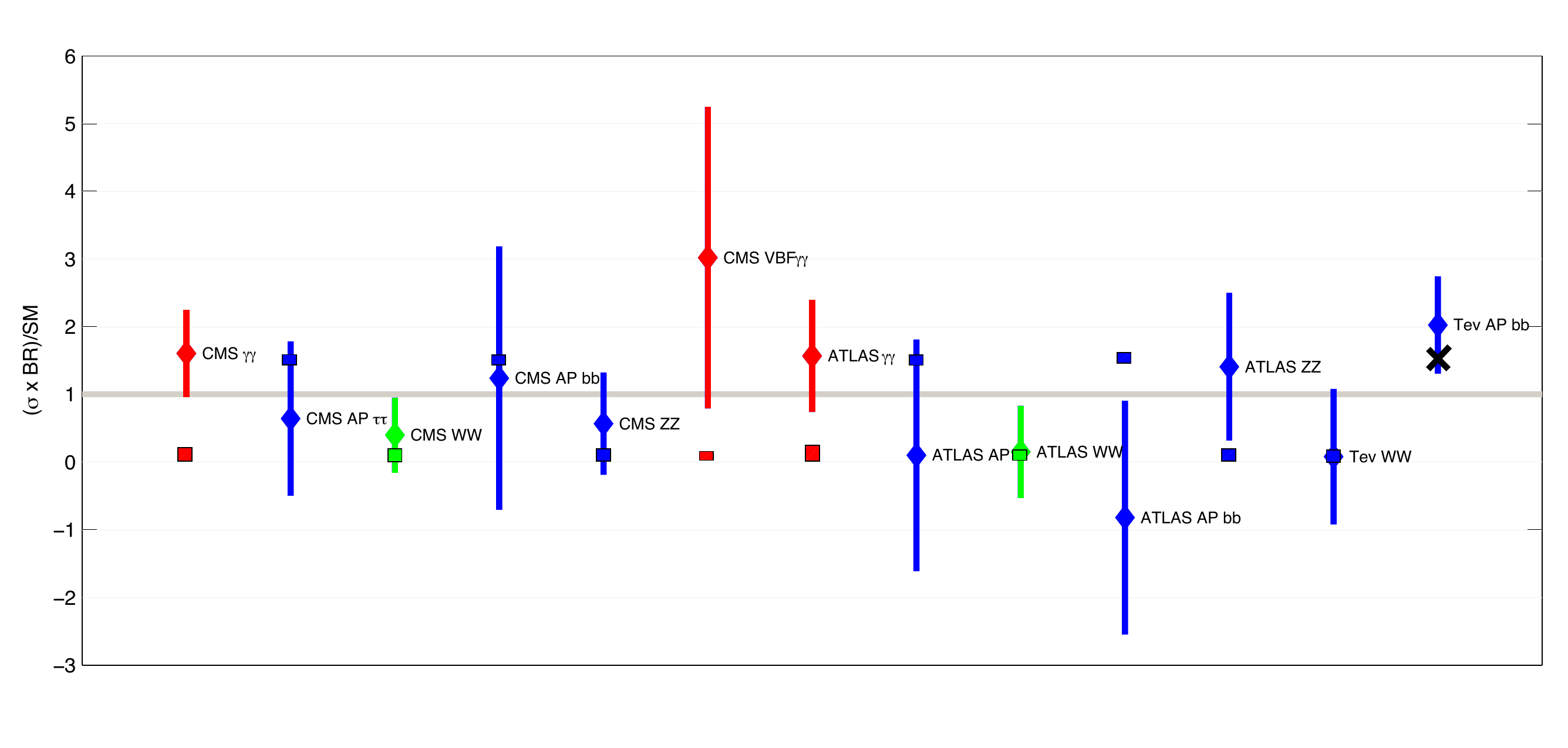} \end{center}
\caption{Natural SUSY predictions, assuming associated production $h\to b\bar b$ as indicated by the black cross, are marked by thick bands.}
\label{fig:ratesBB}
\end{figure}

Second, we select as reference the intriguing Tevatron hint for enhanced $h\to b\bar b$, assuming $\left(\mu_{bb;AP}\right)^{\rm ref}=1.5$. The result is depicted in Fig.~\ref{fig:ratesBB}. In this scenario, natural SUSY predicts a strong suppression in the $VV,\,\gamma\gamma$ final state channels that cannot be compensated by MSSM particle loops, in tension with LHC data.

\section{Conclusions}\label{sec:conc}

We present a detailed study of Higgs couplings in natural SUSY. Our approach is different than most existing analyses in that, while we do keep the discussion centered around the essential recent experimental inputs such as the favored value of the Higgs mass, we do not aim to fit the current data~\cite{Lafaye:2009vr, Klute:2012pu, Giardino:2012ww, Carmi:2012yp, Espinosa:2012ir, Ellis:2012rx, Azatov:2012bz, Azatov:2012rd, Azatov:2012wq}. Instead, our goal is to formulate predictions for Higgs observables in the context of a non-minimal Higgs sector that addresses the little hierarchy problem while staying in accord with $m_h\approx125$ GeV. 
Our framework includes quantum corrections from MSSM particles and allows for an arbitrary 2HDM potential, without restricting to the MSSM structure that is likely distorted by new interactions. 

We separated the analysis into two types of effects: (i) loop corrections, affecting the couplings $h\gamma\gamma$ and $hGG$, and (ii) Higgs mixing, modifying the Higgs-fermion and massive vector boson couplings. 
In Sec.~\ref{sec:loop} we considered loop effects. Demanding fine-tuning no worse than about 1:10, we argued that only stops and charginos can induce coupling modifications larger than $\sim5\%$. Light, unmixed stops enhance $hGG$, the most relevant constraints being electroweak precision tests and direct stop searches. Demanding that $(\Delta\rho/\rho)$ remains within $4\sigma$ from the experimental value constrains the stop correction to  $r_G^{\tilde t}<1.3$. Highly mixed stops can reduce the $hGG$ coupling; here, bounds come from rare $B$ decays and naturalness. Instead of imposing the bound due to $BR(B\to X_s \gamma)$ directly, we incorporate $BR(B\to X_s \gamma)$ into our assessment of the model fine-tuning. This measure disfavors light, mixed stops, with $(X_t\mu\tan\beta/m_{\tilde t}^2) \gsim 1$. We find that $r_G^{\tilde t}>0.9$ is preferred for naturalness. 
Light charginos can vary the $h\gamma\gamma$ vertex in the range $0.7<r_\gamma^{\tilde\chi^\pm}<1.1$, when both states lie close to the direct limit of $\sim$100 GeV. The effect scales as $\propto(1/\tan\beta)$ and goes below $\pm5\%$ for $\tan\beta\gsim5$. 

In Sec.~\ref{sec:hmix} we considered Higgs mixing. We showed that once one assumes that the weak-scale Higgs sector is an approximately type-II 2HDM, with natural flavor conservation broken only by loop effects, then, to a good accuracy, the analysis of Higgs observables depends on only four free parameters, two describing Higgs mixing and two describing loop effects. 
Because of the large branching fraction of a 125 GeV SM-like Higgs to $b\bar b$, the $hb\bar b$ coupling $r_b$ is of key phenomenological importance.
We estimated the size, and interpreted the implications of possible variations in $r_b$, expected within concrete SUSY models. Non-decoupling $D$-term models generically predict $r_b>1$, provided that the MSSM Higgs doublets are taken to transform in a vector representation of the new gauge group. The precise result depends on the mass of the heavier Higgs doublet and can be estimated as $r_b\sim1+2(m_h/m_H)^2$. For $m_H=350$ GeV, consistent with all other experimental constraints, this gives $r_b\sim1.25$ that translates into a $\sim30\%$ reduction in $BR(h\to \gamma\gamma)$, already in some tension with current experimental results. Non-decoupling $F$-term models, consisting of new chiral superfields with hypercharge zero and $\sim$TeV SUSY breaking mass, predict $r_b<1$. 

In Sec.~\ref{sec:data} we listed natural SUSY predictions to Higgs observables and interpreted the current data. 
In particular, we showed that: 
\begin{itemize}
\item Significant enhancement of the $h\to\gamma\gamma$ rate, up to a factor $\sim4$ times the SM result, is viable;
\item The enhancement in $h\to\gamma\gamma$ cannot exceed the enhancement in $h\to ZZ,WW$ by more than 40\%, unless the GF $h\to\gamma\gamma$ rate is itself reduced by more than 25\% compared to the SM;
\item  $h\to b\bar b$ in the associated production channel cannot be enhanced over the SM prediction by more than 50\%.
\end{itemize} 
Our analysis captures the phenomenologically relevant Higgs couplings in natural SUSY models, using four free parameters with restricted numerical ranges. Considering that Higgs studies at the LHC and Tevatron yield $\mathcal{O}(10)$ experimentally independent observables, natural SUSY provides a predictive, falsifiable set-up, with promising opportunities to discriminate between model sub-classes in the near future.

Note added: while this paper was being prepared for submission, a related analysis appeared~\cite{Azatov:2012wq} that partially overlaps with our discussion in Secs.~\ref{sec:Dterm} and~\ref{sec:Fterm}.

\acknowledgments{We thank Nima Arkani-Hamed, Sven Heinemeyer, Yonit Hochberg, Eric Kuflik, Mariangela Lisanti and Nathan Seiberg for useful discussions, and Yossi Nir for comments on the manuscript. KB is supported by DOE grant DEFG0290ER40542. JF is supported by DOE grant DEFG0291ER40671.}

\begin{appendix}
\section{More details of $D$-term models}\label{app:D}

Discussions of $D$-term models can be found e.g. in~\cite{Maloney:2004rc,Dine:2007xi,Carena:2009gx}. Here, for completeness, we repeat the derivation of the low-energy effective potential discussed in Sec.~\ref{sec:Dterm}. Then we comment on the implications of Eq.~(\ref{eq:D05}) for this class of models. 

The link fields ($\Sigma$, $\tilde{\Sigma}$) transform as bi-fundamentals under the product gauge group with a canonical Kahler potential
\beq
K \supset {\rm Tr} e^{g_AV_A} \Sigma  e^{-g_BV_B}  \Sigma^\dagger + {\rm Tr} e^{g_BV_B} \tilde{\Sigma}  e^{-g_AV_A}  \tilde{\Sigma} ^\dagger,
\eeq
where $g_{A,B}$ are the gauge couplings of gauge group $A$ and $B$ and $V_{A,B}$ are the corresponding vector multiplets. After $\Sigma$, $\tilde{\Sigma}$ develop a VEV, $f$, the product gauge group is broken down to the diagonal group, which is identified as the SM electroweak gauge group $G$ with a massless vector multiplet 
\beq
V_{G}=\frac{g_AV_B+g_BV_A}{\sqrt{g_A^2+g_B^2}}.
\eeq
The SM gauge coupling is given by $g_{G}^{-2}=g_A^{-2}+g_B^{-2}$.
The orthogonal combination, 
\beq
V_{H}=\frac{-g_AV_A+g_BV_B}{\sqrt{g_A^2+g_B^2}},
\eeq
is massive, with a mass term in the Kahler potential
\beq
K_V=M_V^2 V_H^2+\cdots,
\eeq
where $M_V=(g_A^2+g_B^2)f^2$. Both MSSM Higgs fields are only charged under $SU(N)_A$ with an initially canonical Kahler potential 
\beq
K_H=\sum_{i=u,d}H_i^\dagger e^{g_AV_A}H_i=\sum_{i=u,d}\left(H_i^\dagger e^{g_G V_G} H_i-\frac{g_{SM}g_A}{g_B}H_i^\dagger V_H H_i\right),
\eeq
where we expand to the leading order in $V_H$. After integrating out $V_H$ through its equation of motion, we have
\beq
K_H^{eff}=\sum_{i=u,d}H_i^\dagger e^{g_G V_G}H_i-\sum_G\frac{g_G^2g_A^2}{g_B^2M_V^2}\left|\sum_{i=u,d}H_i^\dagger T_G^a H_i\right|^2,
\eeq
whose contributions to the Higgs quartic couplings are of order $\mu^2/M_V^2$. Electroweak precision tests require that $M_V > $ 3 TeV~\cite{Chivukula:2003wj}. Given the naturalness condition $\mu<300$ GeV, these supersymmetric corrections are thus too small to explain a Higgs mass at $125$ GeV. 

Introducing SUSY breaking effects via a universal soft mass, $M_s$, to the link fields, modifies the Kahler potential of the heavy vector multiplet,
\beq
K_V=(M_V^2+\theta^4 M_s^2) V_H^2+\cdots.
\eeq
Again after integrating out $V_H$, we have 
\beq
K_H^{eff}=\sum_{i=u,d}H_i^\dagger e^{g_G V_G} H_i-\sum_G\frac{g_G^2g_A^2}{g_B^2}\left(\frac{1}{M_V^2}-\frac{\theta^4M_s^2}{M_V^2+M_s^2}\right)\left|\sum_{i=u,d}H_i^\dagger T_G^a H_i\right|^2,
\label{eq:Kahler2}
\eeq
which gives rise to the low-energy potential in Eq.~(\ref{eq:D-potential}). 

Consider two limits of the modified Higgs couplings in Eq.~(\ref{eq:D-potential}):
\begin{itemize}
\item{$M_V \gg M_s$, corresponding to a nearly supersymmetric heavy threshold, with \beq \Delta (\Delta^\prime)=\frac{g_A^2}{g_B^2}\frac{M_s^2}{M_V^2}\ll\frac{g_A^2}{g_B^2}\nonumber \eeq}
\item{$M_s \gg M_V$, corresponding to hard SUSY breaking in the low energy potential, with \beq \Delta(\Delta^\prime)=\frac{g_A^2}{g_B^2}\nonumber.\eeq}
\end{itemize}
Given that $g_{G}^{-2}=g_A^{-2}+g_B^{-2}$, we cannot take $g_B$ very small at the weak scale. Assuming $g_A/g_B=\mathcal{O}(1)$, we learn that $D$-term models need a SUSY breaking mass $M_s$ at least comparable to the supersymmetric scale $M_V$. These models are thus unlikely to be well described by the nearly supersymmetric effective field theory of~\cite{Dine:2007xi}.

Finally we comment on the fine-tuning in these models. The new non-supersymmetric contribution to the Higgs quartic couplings leads to a quadratic divergence in the Higgs mass squared, cut off by $M_s$. If we require this additional fine-tuning to be no worse than 10\%, 
\beq
\frac{g_G^2\Delta}{16\pi^2}\frac{M_s^2}{m_h^2} < 10,
\eeq
we obtain a conservative upper bound, $M_s \lsim 10$ TeV. Taking into account the electroweak constraint, $M_V > 3$ TeV, and the requirement $M_V < M_s$ coming from Eq.~(\ref{eq:D05}), we find that $M_s$ is limited to the range 3 -- 10 TeV.

\end{appendix}

\bibliography{ref}
\bibliographystyle{jhep}

\end{document}